\documentclass[iop]{emulateapj-rtx4}
\usepackage{apjfonts}
\usepackage{lscape}
\usepackage[usenames]{color}

\newcommand{\kms}{km\,s$^{-1}$} 

\newcommand{\Msun}{$M_{\odot}$}

\newcommand{\rc}{r_{\rm c}}

\newcommand{\cM}{{\cal{M}}}

\newcommand{\mybibitem}[3]{\bibitem[{#1}({#2})]{#3}}
\newcommand{\mybibthree}[4]{\bibitem[{#2}({#3}){#1}]{#4}}
\newcommand{\HST}{\emph{HST} }

\newcommand{\picplace}[1]{\vbox{\hrule\@height 0.4pt\@width\hsize
\hbox to\hsize{\vrule\@width 0.4pt\@height#1\hfil
\vrule\@width 0.4pt\@height#1}\hrule\@height 0.4pt\@width\hsize}}

\slugcomment{Published in ApJ, 797, 35 (2014)}

\shorttitle{Demography of eMSTOs in Intermediate-Age Star Clusters}
\shortauthors{Goudfrooij et al.}

\begin{document}


\title{Extended Main Sequence Turnoffs in Intermediate-Age Star Clusters: \\
 A Correlation Between Turnoff Width and Early Escape Velocity\altaffilmark{1}}


\definecolor{MyBlue}{rgb}{0.3,0.3,1.0}

\author{Paul Goudfrooij$^2$, L\'eo Girardi$^3$, Vera Kozhurina-Platais$^2$, 
  Jason S. Kalirai$^{2,4}$, Imants Platais$^5$, Thomas H. Puzia$^6$, 
  Matteo Correnti$^2$, Alessandro Bressan$^7$, Rupali Chandar$^8$,
  Leandro Kerber$^9$, Paola Marigo$^{10}$, and Stefano Rubele$^3$ \vspace*{0.3mm}}
\affil{$^2$ Space Telescope Science Institute, 3700 San Martin
  Drive, Baltimore, MD 21218, USA; {\color{MyBlue}goudfroo, verap,
    jkalirai, correnti@stsci.edu} \\ 
  $^3$ Osservatorio Astronomico di Padova -- INAF, Vicolo dell'Osservatorio 5,
  I-35122 Padova, Italy; {\color{MyBlue}leo.girardi@oapd.inaf.it} \\
 $^{4}$ Center for Astrophysical Sciences, Johns Hopkins University, 3400
  N. Charles Street, Baltimore, MD 21218, USA \\
 $^{5}$ Department of Physics and Astronomy, Johns Hopkins University, 3400
  N. Charles Street, Baltimore, MD 21218, USA; {\color{MyBlue}imants@pha.jhu.edu} \\
 $^{6}$ Institute of Astrophysics, Pontificia Universidad Cat\'{o}lica de Chile,
  Avenida Vicu\~{n}a Mackenna 4860, Macul, Santiago, Chile;
  {\color{MyBlue}tpuzia@astro.puc.cl} \\ 
 $^{7}$ SISSA, via Bonomea 365, I-34136 Trieste, Italy;
  {\color{MyBlue}alessandro.bressan@sissa.it} \\
 $^{8}$ Department of Physics and Astronomy, The University of Toledo, 2801 West
  Bancroft Street, Toledo, OH 43606, USA; {\color{MyBlue}rupali.chandar@utoledo.edu} \\
 $^{9}$ Universidade Estadual de Santa Cruz, Rodovia Ilh\'eus-Itabuna, km 16, 
  45662-000 Ilh\'eus, Bahia, Brazil; {\color{MyBlue}lkerber@gmail.com}  \\
 $^{10}$ Dipartimento di Fisica e Astronomia Galileo Galilei, Universit\`a di
  Padova, Vicolo dell'Osservatorio 3, I-35122 Padova, Italy;
  {\color{MyBlue}paola.marigo@unipd.it}} 

%\author{Paul Goudfrooij\altaffilmark{2}, Vera Kozhurina-Platais\altaffilmark{2},
%  L\'eo Girardi\altaffilmark{3}, Jasonjot S. Kalirai\altaffilmark{2,4}, Imants
%  Platais\altaffilmark{5}, Thomas H. Puzia\altaffilmark{6}, Leandro
%  Kerber\altaffilmark{7}, Stefano Rubele\altaffilmark{3}, Rupali
%  Chandar\altaffilmark{8}, Alessandro Bressan\altaffilmark{9}, and Paola
%  Marigo\altaffilmark{10}} 
  
\altaffiltext{1}{Based on observations with the NASA/ESA {\it Hubble
    Space Telescope}, obtained at the Space Telescope Science
  Institute, which is operated by the Association of Universities for
  Research in Astronomy, Inc., under NASA contract NAS5-26555} 

\begin{abstract}
We present color-magnitude diagram %(CMD) 
analysis of deep Hubble Space Telescope 
imaging of a mass-limited sample of 18 intermediate-age (1\,--\,2 Gyr old)
star clusters in the Magellanic Clouds, including 8 clusters for which new
data was obtained.   
We find that \emph{all} star clusters in our sample 
feature extended main sequence turnoff (eMSTO) 
regions that are wider than can be accounted for by a simple stellar
population (including unresolved binary stars). 
FWHM widths of the MSTOs
indicate age spreads of 200\,--\,550 Myr. 
We evaluate dynamical evolution of clusters with and without
initial mass segregation.  
Our main results are: 
(1) 
the fraction of red clump (RC) stars in secondary RCs in eMSTO clusters scales with
the fraction of MSTO stars having pseudo-ages $\la$\,1.35 Gyr; 
(2) the width of the pseudo-age distributions of eMSTO clusters is
correlated with their central escape velocity $v_{\rm esc}$, 
both currently and at an age 
of 10 Myr. We find that these two results are unlikely to be reproduced
by the effects of interactive binary stars or a range of stellar rotation
velocities.  
We therefore argue that the eMSTO phenomenon 
is mainly caused by extended star formation within the clusters;
(3) we find that $v_{\rm esc} \geq 15$ \kms\ out to ages of at least 100 Myr for
\emph{all} clusters featuring eMSTOs, while  
$v_{\rm esc} \leq 12$ \kms\ at all ages for two lower-mass clusters in the same
age range that do \emph{not} show eMSTOs. 
We argue that eMSTOs only occur for clusters whose early escape velocities are 
higher than the wind velocities of stars that provide 
material from which second-generation stars can form. 
The threshold of 12\,--\,15 \kms\ is consistent with wind
velocities of intermediate-mass AGB stars and massive binary stars in the
literature.  
\end{abstract}





%----------------------------- intro section ------------------------------

\section{Introduction}              \label{s:intro}

For almost a century and counting, the study of globular clusters (GCs) has
contributed enormously to our understanding of stellar evolution. Until
recently, this was especially true because they were thought to be simple objects
consisting of thousands to millions of coeval stars with the same chemical
composition. However, this notion has had to face serious challenges over the
last $\sim$\,dozen years. It is now commonly recognized that GCs typically harbor
multiple stellar populations featuring several unexpected characteristics
\citep[for recent reviews, see][]{renz08,grat+12}. 

Recent spectroscopic surveys established that light elements like C, N, O,
Na, Mg, and Al show large star-to-star abundance variations (often dubbed
``Na-O anticorrelations'') within virtually all Galactic GCs studied to
date in sufficient detail \citep[][and references therein]{carr+10}. 
These abundance variations have been found among both red giant branch (RGB)
stars and main sequence (MS) stars in several GCs \citep{grat+04}. This clarified
that the variations cannot be due to internal mixing within stars evolving
along the RGB. Instead, their origin must be primordial, being imprinted on the
stars during their formation process. 
The chemical processes involved in causing the light-element abundance
variations have largely been identified as proton capture reactions at $T
\ga 2\times 10^7$ K, such as the CNO and NeNa cycles. 
Currently, the leading candidates for ``polluter'' sources are stars in which
such reactions occur readily and which feature slow stellar winds so
that their ejecta are relatively easy to retain within the potential well of massive
clusters: \emph{(i)} 
intermediate-mass AGB stars ($4 \la {\cal{M}}/M_{\odot} \la 8$, hereafter
IM-AGB; e.g., \citealt{danven07} and references therein), \emph{(ii)} 
rapidly rotating massive stars (often referred to as ``FRMS'';
\citealt{decr+07}) and \emph{(iii)} massive binary stars \citep{demi+09}. 

In the two currently favored formation scenarios, the abundance variations are
due to stars having either formed from or polluted by gas that is a mixture of 
pristine material and material shed by such ``polluters''.  In the ``in situ
star formation'' scenario \citep[see, e.g.,][]{derc+08,derc+10,conspe11}, the
abundance variations are due to a second generation of stars that formed out of gas
clouds that were polluted by winds of first-generation stars to varying extents,
during a period spanning up to a few hundreds of Myr, depending on
the nature of the polluters. In the alternative ``early disc accretion'' scenario
\citep{bast+13}, the polluted gas is instead accreted by low-mass
pre-main-sequence stars during the first $\approx$\,20 Myr after the formation
of the star cluster. Note that in the latter scenario, the  chemical enrichment
that causes the abundance variations currently seen among RGB and MS stars in
ancient GCs would only have occurred by FRMS and massive binary stars, given
the time scales involved.  

An unfortunate issue in distinguishing between these two distinct scenarios for the
formation of GCs is the ancient age of Galactic GCs ($\sim$\,12\,--\,13 Gyr),
which prevents a direct measurement of the short time scales (and hence the
types of stars) involved in the chemical enrichment of the ``polluted'' stars. 

In the context of the nature of Na-O anticorrelations in Galactic GCs, the
recent discovery of extended main sequence turn-offs (hereafter eMSTOs) in
inter\-me\-diate-age (1\,--\,2 Gyr old) star clusters in the Magellanic Clouds
\citep{mack+08a,glat+08,milo+09,goud+09} has generated much interest in the
literature, especially since many investigations concluded that
the simplest viable interpretation of the eMSTOs is the presence of multiple
stellar populations spanning an age interval of several $10^8$ yr within these
clusters \citep[see
also][]{rube+10,rube+11,goud+11a,goud+11b,conspe11,kell+11,mack+13}.    
However, the eMSTO phenomenon has been interpreted in two other main ways: 
spreads in rotation velocity among turnoff stars (hereafter the ``stellar
rotation'' scenario:\ \citealt{basdem09,li+12,yang+13}, but see
\citealt{gira+11}), and a photometric feature of interacting 
binaries within a simple stellar population \citep[the ``interacting binaries'' 
scenario:][]{yang+11,li+12}.  

One avenue to resolving the nature of eMSTOs in intermediate-age star clusters
is to study features of MSTOs that are likely to be caused by differences in the
clusters' dynamical properties and history of mass loss. In 
particular, \citet{goud+11b} studied a sample of 7 intermediate-age clusters
and found that the stars in the ``bright half'' of the eMSTO region on the
color-magnitude diagram (i.e., the ``youngest half'' if the width of the MSTO
is due to a range of ages) showed a significantly more centrally concentrated
radial distribution than the ``faint half'' if the cluster in question had the
following estimated dynamical properties at an age of 10 Myr: \emph{(i)} a
half-mass relaxation time of at least half the current cluster age and
\emph{(ii)} an escape velocity of $\ga$\,15 \kms, similar to observed wind
speeds of intermediate-mass AGB stars \citep{vaswoo93,mars+04}.  
While such differences in radial distributions are consistent with the
``in situ star formation'' scenario, they seem harder to explain 
by 
the stellar rotation scenario. Specifically, it is difficult to understand why
the inner stars in such clusters would have systematically lower rotation
velocities than stars in the outer regions. As to the interacting binaries
scenario, the data available to date does not show any relation between the
binary fractions of clusters with eMSTOs versus those without, or between
clusters with different dynamical properties. Moreover, the brighter / bluer
half of the eMSTOs, which in this scenario is caused by interacting binaries, is
often the most populated part of the MSTO. This is hard to understand in this
context, since interacting binaries are expected to constitute just a very minor
fraction of the stars in these clusters \citep[see also][]{gira+13,yang+13}. 

In an effort to improve the statistics on the presence and demography of eMSTOs
in inter\-me\-di\-ate-age star clusters and to further study potential effects of
dynamical properties on the morphology of MSTOs, we present MSTO properties of
20 intermediate-age star clusters in the Magellanic Clouds in this paper. This
includes 8 such clusters for which new imaging data was obtained using the
Wide Field Camera \#3 (WFC3) on board the \emph{Hubble Space Telescope (HST)}.  
 
This paper is set up as follows. 
Sect.\ \ref{s:sample} describes the data used in this paper and the star cluster
sample. Isochrone fitting is described in Sect.~\ref{s:isofits}. Dynamical
properties and dynamical evolution of the star clusters in our sample are
discussed in Sect.~\ref{s:dynamics}. Sect.~\ref{s:agedist} describes pseudo-age
distributions of the sample star clusters as derived from the MSTO morphology
and presents a correlation between the MSTO widths and the escape velocities of
the clusters at early times. 
Sect.~\ref{s:disc} discusses our findings in the context of predictions of
currently popular scenarios on the nature of eMSTOs, and Sect.~\ref{s:conc}
summarizes our conclusions. 

\section{Cluster Sample and New Data} \label{s:sample}

The selection procedure for our ``full'' target cluster sample is based on
integrated-light photometry in the literature: Clusters are selected to have a 
``S parameter'' \citep{gira+95,pess+08} in the range 35\,--\,40 along
with an unreddened integrated-light $V$-band magnitude $\leq 12.5$. These 
criteria translate to cluster ages between roughly 1.0 and 2.0 Gyr and
masses $\ga 3 \times 10^{4} \; M_{\odot}$. 
The global properties of the star clusters in our ``full'' sample are listed in
Table~\ref{t:sample}.  
At the onset of this study, data of adequate quality was already available in
the \HST archive for several star clusters in this sample. This includes
star clusters in the \HST programs GO-9891 (PI: G. Gilmore; clusters NGC 1852 and
NGC 2154), GO-10396 (PI: J. Gallagher; cluster NGC 419), and GO-10595 (PI:
P. Goudfrooij; clusters NGC 1751, NGC 1783, NGC 1806, NGC 1846, NGC 1987, 
NGC 2108, and LW 431). These data typically consist of images with the Wide
Field Channel of the ACS camera in the F435W, F555W, and/or F814W
filters. Analyses of these data have been published 
before \citep{glat+08,mack+08a,milo+09,goud+09,goud+11a,goud+11b}; here we use
results on those clusters for correlation studies in Sections \ref{s:agedist}
and \ref{s:disc}. The ACS images of clusters NGC 419, NGC 1852, and NGC 2154
were downloaded from the \HST archive and processed as described in
\citet{goud+09}.  

% Place Table 1 here
\begin{table*}[tbh]
\scriptsize
\begin{center}
\caption[]{Global properties of star clusters in our full sample.}
\label{t:sample}
%\begin{tabular*}{13.65cm}{@{}lc@{~~}c@{~~~}c@{~~~}cccccc@{~~~~}c@{}} \tableline \tableline
\begin{tabular}{@{}lc@{~~}c@{~~~}c@{~~~}cccccc@{~~~~}c@{}} \tableline \tableline
\multicolumn{3}{c}{~~} \\ [-2ex]  
\multicolumn{1}{c}{Name} & $V$ mag & Aper. & Ref. & $r_c$ & $r_{\rm eff}$ & 
 Age & [$Z$/H] & $A_V$ & $(m-M)_0$ & Ref. \\ [0.2ex]
% $\log {\cal{M}}_{\rm cl}$ & $v_{\rm esc}$ & $v_{\rm esc,\,7}^{p}$ \\ 
\multicolumn{1}{c}{(1)} & (2) & (3) & (4) & (5) & (6) & (7) & (8) & (9) & (10) & 
 (11) \\ [0.5ex] \tableline 
\multicolumn{3}{c}{~~} \\ [-1.5ex] 
NGC 411  & $11.81 \pm 0.07$ &   50 & 1 & $ 4.23 \pm 0.26$ &  $ 6.12 \pm 0.79$ & 
 $1.45 \pm 0.05$ & $-0.7 \pm 0.1$ & $0.16 \pm 0.02$ & $18.82 \pm 0.03$ & 1 \\
NGC 419  & $10.30 \pm 0.16$ &   50 & 1 & $ 5.48 \pm 2.01$ &  $ 7.67 \pm 2.86$ & 
 $1.45 \pm 0.05$ & $-0.7 \pm 0.1$ & $0.15 \pm 0.02$ & $18.85 \pm 0.03$ & 1 \\
NGC 1651 & $12.13 \pm 0.12$ &   50 & 1 & $ 4.57 \pm 0.36$ &  \llap{1}$2.82 \pm 2.01$ & 
 $2.00 \pm 0.05$ & $-0.3 \pm 0.1$ & $0.19 \pm 0.02$ & $18.41 \pm 0.03$ & 1 \\
NGC 1718 & $12.25 \pm 0.15$ &   31 & 2 & $ 3.74 \pm 0.24$ &  $ 5.42 \pm 0.56$ & 
 $1.80 \pm 0.05$ & $-0.3 \pm 0.1$ & $0.58 \pm 0.03$ & $18.42 \pm 0.03$ & 1 \\
NGC 1751 & $11.67 \pm 0.13$ &   50 & 1 & $ 5.76 \pm 0.41$ &  $ 7.10 \pm 0.87$ & 
 $1.40 \pm 0.05$ & $-0.3 \pm 0.1$ & $0.38 \pm 0.02$ & $18.50 \pm 0.03$ & 2 \\
NGC 1783 & $10.39 \pm 0.03$ &   50 & 1 & \llap{1}$0.50 \pm 0.49$ &  \llap{1}$1.40 \pm 2.24$ & 
 $1.70 \pm 0.05$ & $-0.3 \pm 0.1$ & $0.00 \pm 0.02$ & $18.49 \pm 0.03$ & 2 \\
NGC 1806 & $11.00 \pm 0.05$ &   50 & 1 & $ 5.91 \pm 0.27$ &  $ 9.04 \pm 1.24$ & 
 $1.60 \pm 0.05$ & $-0.3 \pm 0.1$ & $0.05 \pm 0.03$ & $18.50 \pm 0.03$ & 2 \\
NGC 1846 & $10.68 \pm 0.20$ &   50 & 1 & $ 8.02 \pm 0.49$ &  $ 8.82 \pm 0.68$ & 
 $1.70 \pm 0.05$ & $-0.3 \pm 0.1$ & $0.07 \pm 0.02$ & $18.42 \pm 0.03$ & 2 \\
NGC 1852 & $12.01 \pm 0.15$ &   36 & 2 & $ 5.10 \pm 0.46$ &  $ 6.97 \pm 0.83$ & 
 $1.40 \pm 0.05$ & $-0.3 \pm 0.1$ & $0.12 \pm 0.02$ & $18.55 \pm 0.03$ & 1 \\
%NGC 1978 & $10.20 \pm 0.02$ &   50 & 1 & $ 7.51 \pm 0.22$ &  $ 9.48 \pm 0.95$ & 
% $2.45 \pm 0.05$ & $-0.3 \pm 0.1$ & $0.07 \pm 0.02$ & $18.42 \pm 0.03$ & 1 \\
NGC 1987 & $11.74 \pm 0.09$ &   50 & 1 & $ 4.18 \pm 0.46$ &  \llap{1}$2.78 \pm 3.05$ & 
 $1.10 \pm 0.05$ & $-0.3 \pm 0.1$ & $0.12 \pm 0.02$ & $18.37 \pm 0.03$ & 2 \\
NGC 2108 & $12.32 \pm 0.15$ &   31 & 2 & $ 5.42 \pm 0.27$ &  $ 7.20 \pm 0.76$ & 
 $1.00 \pm 0.05$ & $-0.3 \pm 0.1$ & $0.48 \pm 0.02$ & $18.45 \pm 0.03$ & 2 \\
NGC 2154 & $11.85 \pm 0.13$ &   50 & 1 & $ 4.50 \pm 0.29$ &  $ 5.69 \pm 0.51$ & 
 $1.55 \pm 0.05$ & $-0.3 \pm 0.1$ & $0.01 \pm 0.02$ & $18.45 \pm 0.03$ & 1 \\
NGC 2173 & $12.01 \pm 0.14$ &   50 & 1 & $ 3.53 \pm 0.27$ &  $ 6.30 \pm 1.10$ & 
 $1.55 \pm 0.05$ & $-0.3 \pm 0.1$ & $0.28 \pm 0.02$ & $18.37 \pm 0.03$ & 1 \\
NGC 2203 & $11.29 \pm 0.15$ &   75 & 2 & $ 7.99 \pm 0.39$ &  $ 9.48 \pm 1.58$ & 
 $1.55 \pm 0.05$ & $-0.3 \pm 0.1$ & $0.16 \pm 0.02$ & $18.37 \pm 0.03$ & 1 \\
NGC 2213 & $12.37 \pm 0.10$ &   50 & 1 & $ 2.57 \pm 0.15$ &  $ 3.57 \pm 0.29$ & 
 $1.70 \pm 0.05$ & $-0.3 \pm 0.1$ & $0.14 \pm 0.02$ & $18.36 \pm 0.03$ & 1 \\
LW 431   & $13.67 \pm 0.15$ &   19 & 2 & $ 4.03 \pm 0.24$ &  $ 9.10 \pm 3.16$ & 
 $1.90 \pm 0.05$ & $-0.3 \pm 0.1$ & $0.14 \pm 0.02$ & $18.45 \pm 0.03$ & 2 \\
Hodge 2  & $11.90 \pm 0.15$ &   31 & 2 & $ 2.67 \pm 0.41$ &  $ 9.09 \pm 2.33$ & 
 $1.30 \pm 0.05$ & $-0.3 \pm 0.1$ & $0.15 \pm 0.02$ & $18.40 \pm 0.03$ & 1 \\
Hodge 6  & $12.09 \pm 0.15$ &   50 & 2 & $ 4.47 \pm 0.49$ &  $ 5.54 \pm 0.87$ & 
 $2.25 \pm 0.05$ & $-0.3 \pm 0.1$ & $0.25 \pm 0.02$ & $18.40 \pm 0.03$ & 1 \\ [0.5ex] \tableline
\multicolumn{5}{c}{~~} \\ [-1.2ex]
\end{tabular}                   
\tablecomments{Column (1): name of star cluster. Column (2): integrated $V$-band 
 magnitude. Column (3): radius of aperture used for integrated-light 
 photometry in arcsec. Column (4): reference of integrated-light photometry (1 =
 \citealt{goud+06}; 2 = \citealt{bica+96}). Column (5): core radius in pc. Column
 (6): effective radius in pc. Column (7): (mean) age in Gyr. Column (8): metallicity
 in dex. Column (9): $V$-band foreground extinction. Column (10): distance
 modulus. Column (11): reference of data in columns 5\,--\,10 (1 = this
 paper; 2 = \citealt{goud+11a}). 
}
\end{center}
%\parbox{8.3cm}{
%{\sl Notes}.~~Column (1): NGC number of galaxy. 
%Column (2): absolute $V$ magnitude of galaxy. Column (3:) ${\it col}_0$ of galaxy. 
%Column (4): ${\it col}_0$ of red GCs. Column (5): color gradient $G_{\it col}$ of red GCs.   
%Column (6): mean rms error of fit of Eq.\ (1) to colors of red GCs
%(in mag).}
\end{table*}

For the remaining 8 clusters in our sample, new data were acquired as part of
\HST program GO-12257 (PI: L. Girardi), using the UVIS channel of WFC3. 
Multiple exposures were taken with the F475W and F814W filters. 
The new WFC3 data consists of 2 or 3 long exposures plus one short 
exposure in each filter. The short exposures were taken to avoid saturation of the
brightest stars in the cluster, and are only used for photometry of those bright
stars\footnote{This was done to avoid significant charge transfer inefficiency
  at low source count levels when the sky level is low \citep[see, e.g.,][]{noes+12}.}. 
The long exposures in each filter were spatially offset by several pixels from
one another in order to simplify the identification and removal of bad detector
pixels in the photometric analysis. 
The target clusters were centered on one of the two CCD chips of WFC3 so 
as to cover both the central regions of the clusters and a fairly large radial
extent to reach the field component.  
A journal of the new observations is listed in Table~\ref{t:obs}.

% Place Table 2 here
\begin{table}[tbh]
\begin{center}
\footnotesize
%\scriptsize
\caption{Journal of WFC3 observations of 8 star clusters.}
 \label{t:obs}
\begin{tabular}{@{}lrccc@{}}
\multicolumn{3}{c}{~} \\ [-2.5ex]   
 \tableline \tableline
\multicolumn{3}{c}{~} \\ [-1.8ex]                                                
\multicolumn{1}{c}{Cluster} & \multicolumn{1}{c}{Obs.\ Date} & $t_{\rm exp,\,F475W}$ &
% $t_{\rm exp, F555W}$ & 
 $t_{\rm exp,\,F814W}$   \\
\multicolumn{1}{c}{(1)}     & \multicolumn{1}{c}{(2)}        & (3) & % (4) & 
 (4) \\ [0.5ex] \tableline  
\multicolumn{3}{c}{~} \\ [-1.5ex]              
%                                  
%NGC 411  & Aug 15, 2011 & 1520 & --- & 1980 \\
%NGC 1651 & Oct 16, 2011 & 1440 & --- & 1520 \\
%NGC 1718 & Dec 02, 2011 & 1440 & --- & 1520  \\
%NGC 1852 & Oct 12, 2011 & ---  & 1040 & ---  \\ 
%%NGC 1978 & Aug 15, 2011 & ---  & 1040 & ---  \\ 
%NGC 2154 & Oct 09, 2011 & ---  & 1040 & --- \\ 
%NGC 2173 & Oct 09, 2011 & 1520 & --- & 1980  \\
%NGC 2203 & Oct 08, 2011 & 1520 & --- & 1980  \\ 
%NGC 2213 & Nov 29, 2011 & 1440 & --- & 1520 \\
%Hodge 2  & Jan 21, 2012 & 1440 & --- & 1520  \\
%Hodge 6  & Aug 16, 2011 & 1440 & --- & 1520  \\ [0.5ex] \tableline
NGC 411  & Aug 15, 2011 & 1520 & 1980 \\
NGC 1651 & Oct 16, 2011 & 1440 & 1520 \\
NGC 1718 & Dec 02, 2011 & 1440 & 1520  \\
NGC 2173 & Oct 09, 2011 & 1520 & 1980  \\
NGC 2203 & Oct 08, 2011 & 1520 & 1980  \\ 
NGC 2213 & Nov 29, 2011 & 1440 & 1520 \\
Hodge 2  & Jan 21, 2012 & 1440 & 1520  \\
Hodge 6  & Aug 16, 2011 & 1440 & 1520  \\ [0.5ex] \tableline
\multicolumn{3}{c}{~} \\ [-2.5ex]              
\end{tabular}
\tablecomments{Column (1): Name of star cluster. (2): Date of {\it
    HST/WFC3\/} observations. (3) Total exposure time in F475W filter in
  seconds. %(4): Total exposure time in F555W filter. 
  (4): total exposure time in F814W filter. }
\end{center}
\end{table}

The data reduction and analysis of the WFC3 data were very similar to
those described in \citet{goud+11a}. Briefly, stellar photometry is  conducted
using point-spread function (PSF) fitting using the spatially-variable
``effective point spread function'' (hereafter ePSF) package developed by J.\
Anderson \citep[e.g.,][]{ande+08} and later adapted by him for use with WFC3 
imaging. This method performs PSF fitting on each individual flat-fielded image  
from the \HST pipeline, using a library of well-exposed PSFs
for the different filters, and adjusting for differing focus among the exposures
(often called ``breathing''). 
We selected all stars with the ePSF parameters ``PSF fit quality''  $q < 0.5$
and ``isolation index'' = 5. The latter parameter selects stars that have no
brighter neighbors within a radius of 5 pixels.  Finally, we 
match the stars detected in all individual images to a tolerance of 0.2 pixel
and perform a weighted combination of the photometry.  

Photometric errors and incompleteness fractions as functions of stellar
brightness, color, and position within the image are quantified by repeatedly
adding small numbers of artificial ePSFs to all individual flat-fielded images of
a given cluster, covering the magnitude and color ranges of stars found in the
color-magnitude diagrams (CMDs), and then re-running the ePSF
software. The overall radial distribution of the artificial stars was
chosen to follow that of the cluster stars (see \S\,\ref{s:kingfits}).  An
inserted star was considered recovered if the input and output 
magnitudes agreed to within 0.75 mag in both filters. Completeness fractions
were assigned to every individual star by fitting the
completeness fractions of artificial stars as functions of their magnitude and
distance from the cluster center.

To check for consistency with other photometry packages, we also analyzed the
data of two star clusters in our sample using P. Stetson's DAOPHOT package as
implemented in the pipeline described in \citet{kali+12}. Both methods
produced consistent results. In the following, we will use the photometry
resulting from the ePSF package. 
The photometry tables of the clusters in our sample, including
completeness fractions, can be requested from the first author. 

\section{Isochrone Fitting} \label{s:isofits}

We derive best-fit ages and metallicities ([Z/H]) of the target clusters 
for which new data was obtained (i.e., new WFC3 data or ACS data from the \HST
archive) using Padova isochrones \citep{mari+08}. Using their web
site\footnote{http://stev.oapd.inaf.it/cgi-bin/cmd}, we construct two grids of
isochrones (one for the HST/WFC3 filter passbands and one for HST/ACS) covering
the ages $0.7 \leq \tau({\rm Gyr}) \leq 2.5$ with a step of 0.05 Gyr and
metallicities $Z$ = 0.002, 0.004, 0.006, 0.008, 0.01, and 0.02. 
Isochrone fitting was performed using methods described 
in \citet{goud+09,goud+11a}. Briefly, we use the observed difference 
in (mean) magnitude between the MSTO and the RC (which is primarily sensitive
to age) along with the slope of the RGB (which is primarily sensitive to
[Z/H]). 
See \citet{goud+11a} for details on how these parameters are
determined. 
We then select all isochrones for which the values of the two parameters
mentioned above lie within 2 $\sigma$ of the measurement uncertainty of those
parameters on the CMDs.  
For this set of roughly 5\,--\,10 isochrones per cluster, we then find the
best-fit values for distance modulus $(m-M)_0$ and foreground reddening $A_V$
by means of a least squares fitting program to the magnitudes and colors of
the MSTO and RC. For the filter-dependent dust extinction we use $A_{\rm
  F475W} = 1.192\, A_V$ and $A_{\rm F814W} = 0.593\, A_V$ for the WFC3
filters, and $A_{\rm F555W} = 1.026\, A_V$ and $A_{\rm F814W} = 0.586\, A_V$ 
for the ACS filters. These values were
derived using the filter passbands in the {\sc synphot} package of 
STSDAS\footnote{STSDAS is a product of the Space Telescope Science Institute,
  which is operated by AURA for NASA.} along with the reddening law of
\citet{card+89}. 
Finally, the isochrones were overplotted onto the CMDs for visual
examination and the visually best-fitting one was selected. Uncertainties of
the various parameters were derived from their variation among the 5\,--\,10
isochrones selected prior to this visual examination \citep[see][for
details]{goud+11a}. 

The best-fit population properties of the clusters are listed in
Table~\ref{t:sample}, along with their integrated $V$-band magnitudes from the
literature. For the clusters that were analyzed before in \citet{goud+11a}, we
only list the properties that resulted from their analysis using the
\citet{mari+08} isochrones for consistency reasons.

\section{Dynamical Properties of the Clusters}  \label{s:dynamics}

If the eMSTO phenomenon is due (at least in part) to a range in stellar ages
in star clusters as in the ``in situ'' scenario, 
  the clusters must have an adequate amount of gas available to form
  second-generation stars. Plausible origins of this gas could be accretion from
  the surrounding interstellar medium (ISM; see, e.g., \citealt{conspe11})
  and/or retention of gas lost by the first generation of stars.  
One would expect the ability of star clusters to retain
the latter material to scale with their escape velocities at the time the
candidate ``polluter'' stars are present in the cluster \citep[see][for
details]{goud+11b}. Conversely, one would not expect to see significant
correlations between the eMSTO morphology and dynamical properties of the
clusters if eMSTOs are mainly due to a range of stellar rotation velocities. 
With this in mind, we estimate masses and escape velocities of the sample
clusters as a function of time going back to an age of 10 Myr, after the
cluster has survived the era of gas expulsion and violent relaxation and when
the most massive stars of the first generation proposed to be candidate
polluters in the literature (i.e., FRMS and massive binary stars) are expected
to start losing significant amounts of mass through slow winds. 

\subsection{Present-Day Masses and Structural Parameters} \label{s:kingfits}

\begin{figure*}[tbhp]
\centerline{\includegraphics[width=12cm]{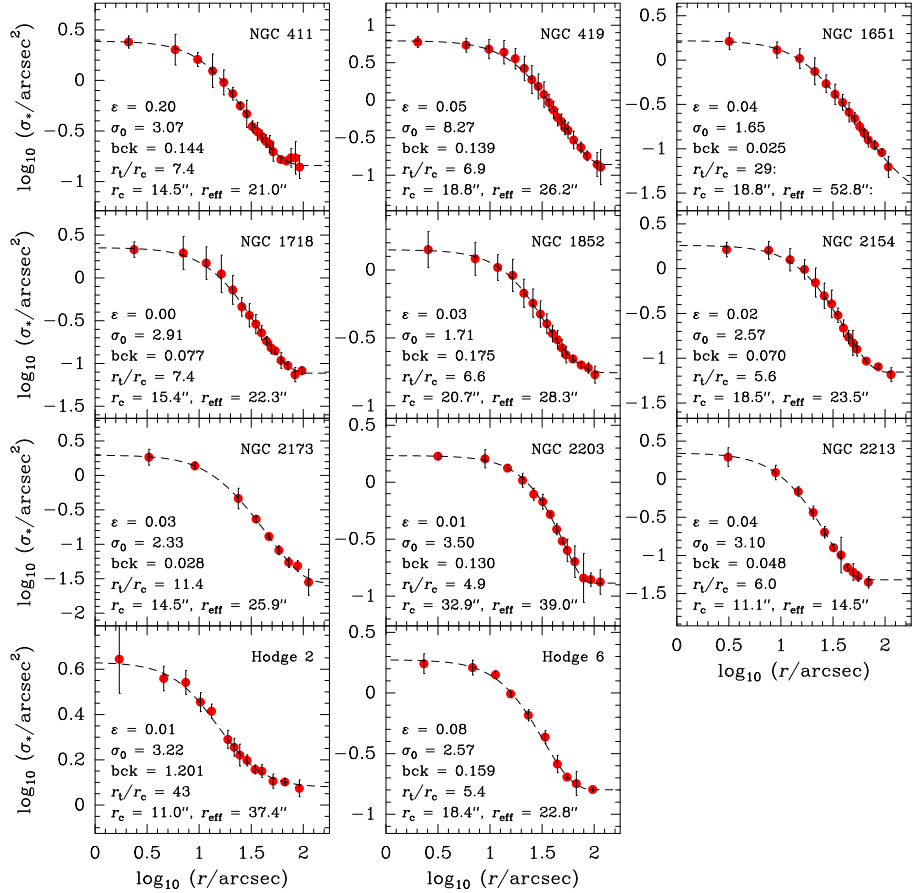}}
%\centerline{\includegraphics[width=0.85\textwidth]{surfdensplot2.ps}}
\caption{Radial surface number density profiles of the star clusters in our
  sample that were not analyzed before in \citet{goud+09} or
  \citet{goud+11a}. The points represent observed (completeness-corrected)
  values, while dashed lines represent the best-fit \citet{king62} models
  whose parameters are shown in the legends, along with the names and
  ellipticities of the clusters.  
}
\label{f:kingfits}
\end{figure*}

Structural parameters of the star clusters in our sample are determined by
fitting elliptical \citet{king62} models to completeness-corrected radial
surface number density profiles, following the method described in \S\,3.3 of 
\citet{goud+11a}. We only use stars brighter than the magnitude at which an
incompleteness of 75\% occurs in the innermost region of the cluster in
question (typically around ${\rm F475W}_{\rm WFC3} \simeq 23.5$ or ${\rm 
  F555W}_{\rm ACS} = 22.8$). Figure~\ref{f:kingfits} shows the best-fit King
models along with the individual surface number density distributions for each
star cluster in our sample (except the clusters for which King model fits were 
performed and illustrated in \citealt{goud+11a}). 

Cluster masses are determined from the $V$-band
magnitudes listed in Table~\ref{t:sample}. Using the measurement aperture
size of those integrated-light magnitudes along with the best-fit King model
parameters of each cluster, we first determine the fraction of total cluster
light encompassed by the measurement aperture. This is done using a
routine that interpolates within points of a fine radial grid while
calculating the integral of the \citet{king62} function. 
After correcting the integrated-light $V$ magnitudes for the missing cluster
light beyond the measurement aperture\footnote{This
  correction was not done by \citet{goud+11b} or 
  \citet{conspe11}.}, total cluster masses are calculated from the
values of $A_V$, $(m-M)_0$, [$Z$/H], and age listed in Table~\ref{t:sample}. 
This process involves interpolation between the ${\cal{M}}/L_V$ values
listed in the SSP models of \citet{bc03}, assuming a \citet{salp55} initial
mass function (IMF)\footnote{For reference, the stellar mass range covered by
  the CMDs of the star clusters studied here is $\approx$\,0.8\,--\,1.9
  $M_{\odot}$. If a \citet{krou01} or \citet{chab03} IMF would be used instead,
  the derived cluster masses would decrease by a factor $\simeq$\,1.6 although all
  mass-related trends among clusters would remain the same.}.  
The latter models were recently found to provide the best fit (among popular
SSP models) to observed integrated-light photometry of LMC clusters with ages and
metallicities measured from CMDs and spectroscopy of individual RGB stars in
the 1\,--\,2 Gyr age range \citep{pess+08}. 

\subsection{Dynamical Evolution of the Star Clusters} \label{s:dynevol}

We perform the dynamical evolution calculations described in \citet{goud+11b}
for all clusters in our sample. Briefly, this involves the evaluation of the
evolution of cluster mass and effective radius for model clusters with and
without initial mass segregation. 
All calculations cover an age range of 10 Myr to 13 Gyr, and take into account
the effects of stellar evolution mass loss and internal two-body
relaxation. For the case of model clusters with initial mass segregation, we
adopt the results of the simulation called SG-R1 in \citet{derc+08}, which
involves a tidally limited model cluster that features a level of initial mass
segregation of $r_{\rm e} / r_{{\rm e,}\,>1} = 1.5$, where $r_{{\rm e,}\,>1}$
is the effective radius of the cluster for stars with ${\cal{M}} > 1\,M_{\odot}$. 
  The primary reason why this simulation was selected for the purposes of the
  current paper is that it yields a number ratio of first-to-second-generation
  stars (hereafter called FG:SG ratio) of $\approx$\,1:2 at an age of $\sim$\,1.5
  Gyr, which is similar to that seen in the clusters in our sample with
  the largest core radii (e.g., NGC 419, NGC 1751, NGC 1783, NGC 1806, NGC
  1846), which likely had the highest levels of initial mass segregation
  \citep[][see discussion in \S\,\ref{sub:SFHs} below]{mack+08b}. In contrast,
  the simulations of clusters that do not fill their Roche lobes by
  \citet[][e.g., their SG-R05, SG-R06, and SG-R075 models]{derc+08} have FG:SG
  ratios $\sim$\,5:1\,--\,3:1, which are inconsistent with the observations of
  the aforementioned clusters in our sample\footnote{The models
    of \citet{derc+08} employed a tidal field strength appropriate to that in
    our Galaxy at a galactocentric distance of 4 kpc. This suggests that the
    tidal field was stronger when the massive clusters in our sample were
    formed than it is now at their current locations, perhaps due to physical 
    conditions prevailing during tidal interactions between the Magellanic
    Clouds 1\,--\,2.5 Gyr ago which caused strong star formation in the bar
    and NW arm of the LMC \citep[e.g.,][]{diabek11,besl+12,rube+12,piat14}.}. 

We note that several young clusters in the Milky Way and the Magellanic Clouds
exhibit high levels of mass segregation which is most likely primordial in 
nature 
  \citep[e.g.,][]{hilhar98,fisc+98,siri+00,siri+01,degr+02,mack+08b}. For
  some clusters the level of mass segregation is actually higher than that in
  the SG-R1 model mentioned above 
(e.g., R136, in which stars with ${\cal{M}} > 3 \; M_{\odot}$ are a
factor $\sim$\,4 more centrally concentrated than stars with ${\cal{M}} < 3 \;
M_{\odot}$, see \citealt{siri+00}). For clusters with such high levels of
initial mass segregation, the simulations of \citet{vesp+09} suggest that they
may dissolve in a few Gyr if they fill their Roche lobe. Hence, one should not 
discard the possibility that some of the intermediate-age clusters in the
Magellanic Clouds may actually dissolve before reaching ``old age''. This fate
may be most likely for the intermediate-age clusters with the largest core
radii and/or the lowest current masses (see discussion in \S\,\ref{sub:correl}
below).    

To estimate the systematic uncertainty of our mass loss rates for the case of
clusters with initial mass segregation, we repeat our calculations for the
case of the SG-C10 simulation of \citet{derc+08}, which yields a FG:SG ratio
at an age of 1.5 Gyr that is somewhat smaller than the SG-R1 simulation, while
still being broadly consistent with the FG:SG ratios observed in the clusters
in our sample with the largest core radii. A comparison of the two
calculations indicates that the systematic uncertainty of our mass loss rates
for the case of initial mass segregation is of order 30\%.

\subsection{Escape Velocities} \label{s:vesc}

Escape velocities are determined for every cluster by assuming a
single-mass King model with a radius-independent $\cM/L$ ratio as calculated
above from the clusters' best-fit age and [$Z$/H] values. Escape velocities
are calculated from the reduced gravitational potential, 
$v_{\rm esc} (r,t) = (2\Phi_{\rm tid} (t) - 2\Phi (r,t))^{1/2}$, at the core
radius\footnote{We acknowledge that this will
  underestimate somewhat the $v_{\rm esc}$ values for clusters with
  significant mass segregation.}. Here
$\Phi_{\rm tid}$ is the potential at the tidal (truncation) radius of the
cluster. 
We choose to calculate escape velocities at the cluster's core radius in view
of the prediction of the ``in situ'' scenario, i.e., that the second-generation
stars are formed in the innermost regions of the cluster
\citep[e.g.,][]{derc+08}. Note that this represents a change relative to
  the escape velocities in \citet{goud+11b}, which were calculated at the 
  effective radius. For reference, the ratio between the two escape velocities
  can be approximated by 
  $v_{{\rm esc},\,r_c}/v_{{\rm esc},\,r_e} = 1.1075 + 0.4548 \, \log c -
  0.4156 \, (\log c)^2 + 0.1772 \, (\log c)^3$ 
  where $c = r_{\rm t}/r_{\rm c}$ is the King concentration parameter. For
  King models with $5 < c < 130$, this approximation is accurate to within
  $\approx$\,3\% rms.   

% Place Table 3 (deluxetable) here. 
% (At the end of the document for the emulateapj version)

For convenience, we define ${\cal{M}}_{\rm cl, 7} \equiv {\cal{M}}_{\rm cl} \, (t =
10^7 {\rm yr})$ and $v_{\rm esc, 7} (r) \equiv v_{\rm esc}\,(r, t = 10^7 {\rm
  yr})$ hereinafter, and refer to them as ``early cluster mass'' and ``early escape 
velocity'', respectively. Masses and escape velocities of the clusters in our
full sample are listed in Table~\ref{t:dynamics} (at the end of this manuscript), 
both for the current ages and for an age of 10 Myr.

\section{Pseudo-Age Distributions} \label{s:agedist}

\subsection{Methodology} \label{sub:method}

\begin{figure*}[tbhp]
%\centerline{\includegraphics[width=0.9\textwidth]{f2.ps}}
\centerline{\includegraphics[width=0.68\textwidth]{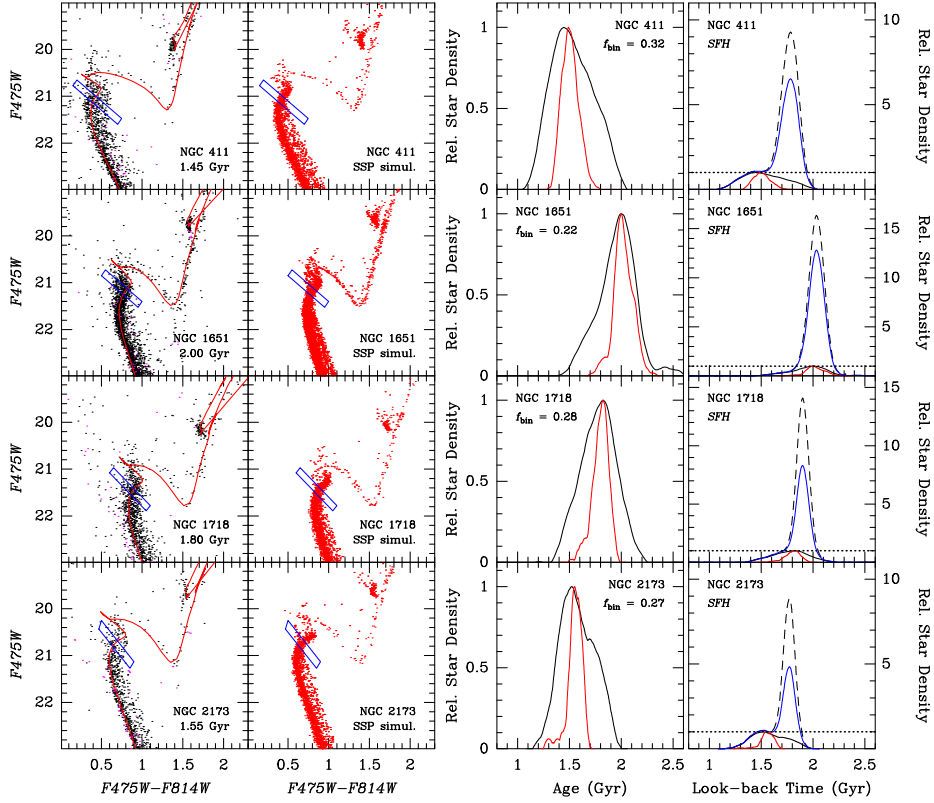}}
\caption{\emph{Left panels}: F475W versus F475W\,$-$\,F814W CMDs of 4 star
  clusters in our sample with new WFC3 observations. Magnitudes are in the
  Vega-based system. Cluster names and ages are mentioned
  in the legends. Black dots represent stars within the
  clusters' core radius. Magenta dots represent stars within ``background
  regions'' far away from the cluster center, with the same total area as the
  area within the core radius. The blue parallelogram depicts the region from
  which pseudo-age distributions were derived (see \S\,\ref{sub:method}).
  \emph{Second column of panels}: F475W versus F475W\,$-$\,F814W CMDs of the simulated
  star clusters (see \S\,\ref{sub:MCsim}). The blue parallelogram from the left
  panels is included for comparison purposes.
  \emph{Third column of panels}: Pseudo-age distributions of the star clusters (black
  lines) and of the associated SSP simulations (red lines). The best-fit binary
  fractions determined during the simulations are reported below the cluster
  names in the legend. 
  \emph{Right panels}: Estimates of star formation histories (SFHs) of the star
    clusters. The black and red lines are the same as in the third row of
    panels. The dashed lines indicate the SFHs of the clusters according to the
    SG-R1 model of \citet{derc+08}. The solid blue lines indicate the clusters'
    SFHs according to an estimated level of initial mass segregation for each
    cluster. The amplitudes of the SFHs are relative to the maximum star density
    reached in the pseudo-age distributions of the respective clusters. See
    \S\,\ref{sub:SFHs} for details.  
}
\label{f:cmdplot1}
\end{figure*}

\begin{figure*}[tbhp]
%\centerline{\includegraphics[width=0.9\textwidth]{f4.ps}}
\centerline{\includegraphics[width=0.68\textwidth]{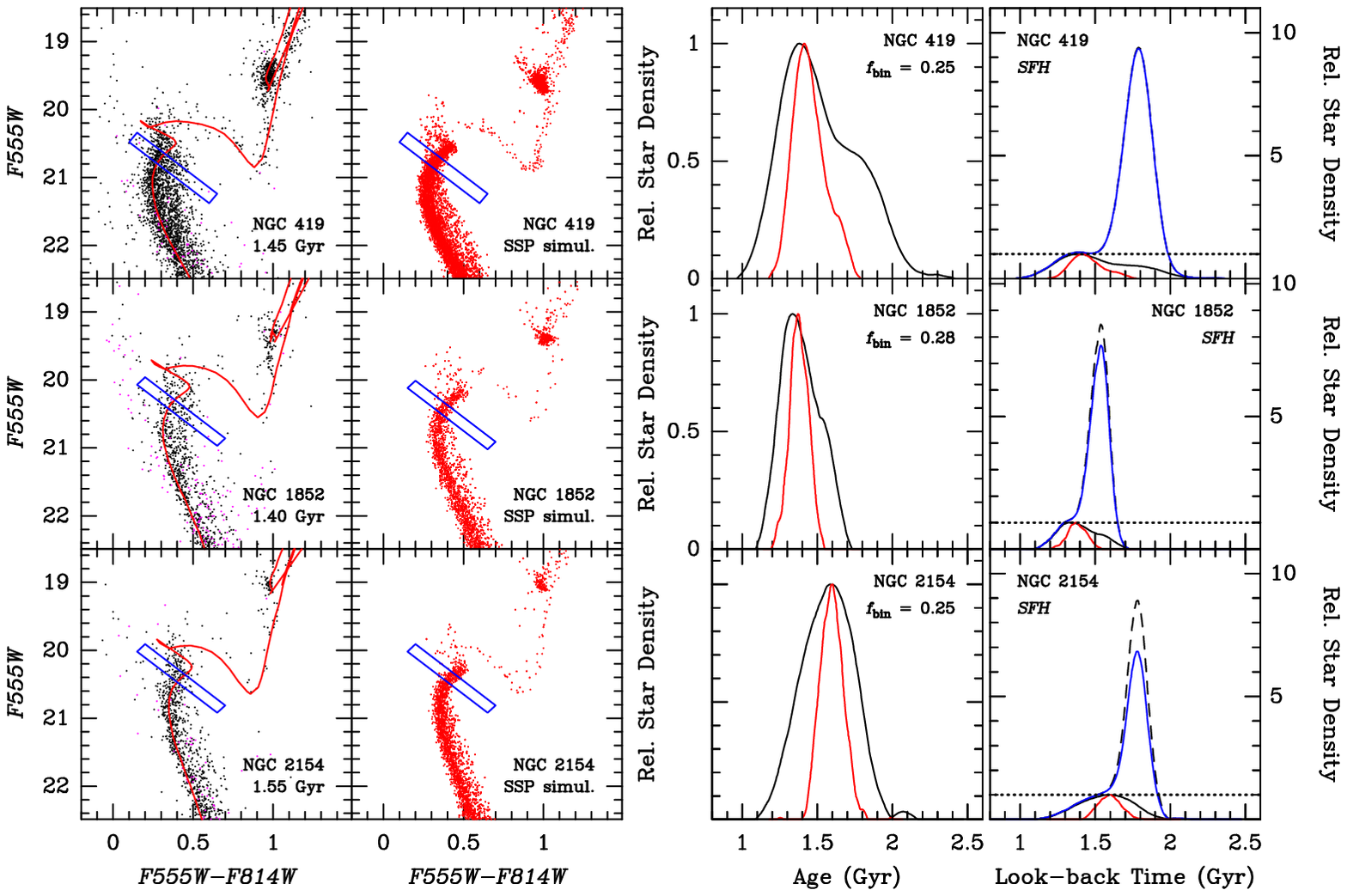}}
\caption{Same as Figure~\ref{f:cmdplot1}, but now F555W versus F555W\,$-$\,F814W
  CMDs for 3 other star clusters in our sample using ACS observations from the
  HST archive. 
}
\label{f:cmdplot3}
\end{figure*}

``Pseudo-age'' distributions of the clusters in our sample are compiled
following the steps  described in Sect.\ 6.1 in \citet{goud+11b}. We construct
a parallelogram in the region of the MSTO where the split between isochrones
of different ages is evident and where the influence of unresolved binary
stars is only minor (see \S\,\ref{sub:MCsim} below and
Figs.~\ref{f:cmdplot1}\,--\,\ref{f:cmdplot3}), just below the `hook' in the
isochrone where core contraction occurs. One axis of the parallelogram is
approximately parallel to the isochrones and the other axis approximately
perpendicular to the isochrones. The magnitudes and colors of stars in the
parallelogram are then transformed into the coordinate frame defined by the two
axes of the  parallelogram, after which we consider the distribution of stars in
the coordinate perpendicular to the isochrones. The latter coordinate is
translated to age by repeating the same procedure for the isochrone tables for
an age range that covers the observed extent of the MSTO region of the cluster
in question (using the same values of [Z/H], $(m\!-\!M)_0$ and $A_V$), and
conducting a polynomial least-squares fit between age and the coordinate
perpendicular to the isochrones.  

The observed ``pseudo-age'' distributions of the clusters are compared to the
distributions that would be expected in case the clusters are true SSPs
(including unresolved binary stars) by conducting Monte-Carlo simulations as
described below. 

\begin{figure*}[tbhp]
%\centerline{\includegraphics[width=0.9\textwidth]{f3.ps}}
\centerline{\includegraphics[width=0.68\textwidth]{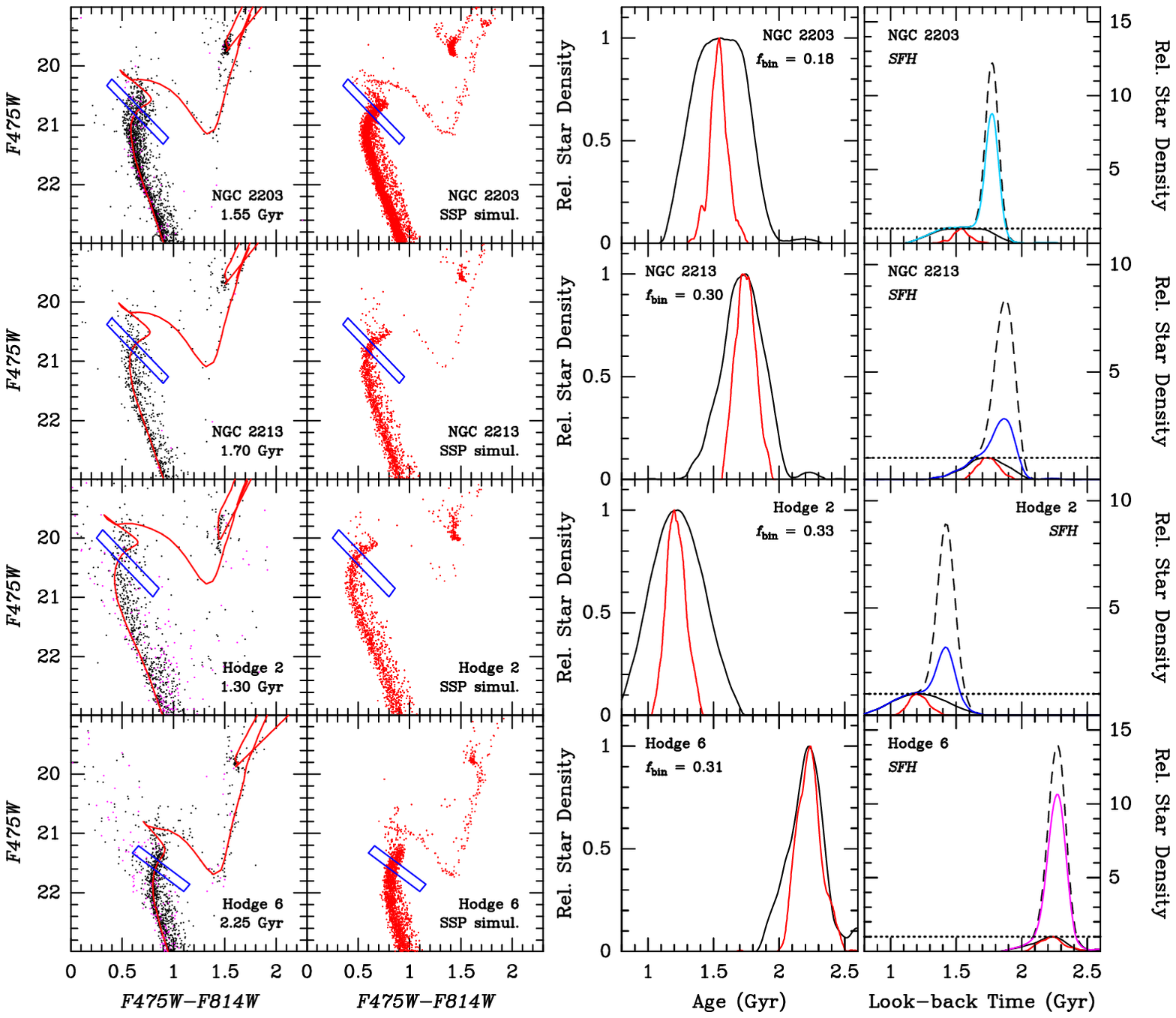}}
\caption{Same as Figure~\ref{f:cmdplot1}, but now for 4 other star clusters
  with new WFC3 observations. The SFHs of NGC 2203 and Hodge 6 are
  plotted with cyan and magenta lines, respectively, for reasons
  discussed in \S\,\ref{sub:SFHs} (below equation \ref{eq:NR_FG}).
}
\label{f:cmdplot2}
\end{figure*}

\subsection{Monte Carlo Simulations} \label{sub:MCsim}

We simulate cluster CMDs of SSPs by populating the best-fit \citet{mari+08} 
isochrones (cf.\ \S\,\ref{s:isofits}) 
with stars randomly drawn from a Salpeter IMF between the minimum
and maximum stellar masses in the isochrone. 
The total number of stars in each simulation is normalized to the 
number of cluster stars on the CMD brighter than the 50\% completeness limit. 
We add unresolved binary companions to a fraction (see below) of the
stars, using a flat primary-to-secondary mass ratio distribution. Finally, we
add random photometric errors to the simulated stars using the
actual distribution of photometric uncertainties established during the 
artificial star tests. 

We use the width of the upper main sequence, i.e. the part brighter than 
the turn-off of the field stellar population and fainter than the MSTO region
of the clusters, to determine the binary star fraction in our sample clusters. 
The latter are mentioned on the right panels of
Figs.~\ref{f:cmdplot1}\,--\,\ref{f:cmdplot3}.    
We estimate the internal systematic uncertainty of the binary fraction as $\pm
5\%$. For the purposes of this work the results don't change significantly
within $\sim\!10\%$ of the binary fraction.  

The pseudo-age distributions of the clusters and their SSP simulations are
depicted in Figs.~\ref{f:cmdplot1}\,--\,\ref{f:cmdplot3}. The left column
of panels show the observed CMDs, the best-fit isochrone (whose properties are
listed in Table~\ref{t:sample}), and the parallelogram mentioned 
in \S\,\ref{sub:method} above (the latter in blue), the second column of panels
show the simulated CMDs along with that same parallelogram, and the 
  third column of panels show the pseudo-age distributions. 
The latter were calculated using the non-parametric Epanechnikov-kernel
probability density function \citep{silv86}, which avoids 
biases that can arise if fixed bin widths are used. In the case of
the observed CMDs, this was done both for stars within the King core radius and
for a ``background region'' far away from the cluster center.  
The intrinsic probability density function of the pseudo-age distribution of
the clusters was then derived by statistical subtraction of the background
regions. Stars in these background regions are plotted on the left panels of 
Figs.~\ref{f:cmdplot1}\,--\,\ref{f:cmdplot3} as magenta dots.
In the case of the simulated CMDs, the pseudo-age distributions are measured
on the average of 10 Monte Carlo realizations. 

\subsection{Fraction of Clusters with eMSTOs in our Sample} 
\label{sub:eMSTOfrac}

As can be appreciated from the third column of panels of
Figs.\ \ref{f:cmdplot1}\,--\,\ref{f:cmdplot3}, the pseudo-age distributions of
\emph{all but one} clusters in our sample are significantly wider than that of
their respective SSP simulations. 
This includes the case of NGC~2173, for which previous studies using
ground-based data rendered the presence of an eMSTO uncertain (see
\citealt{bert+03} versus \citealt{kell+12}). We postulate that the effect of
crowding on ground-based imaging in the inner regions of many star clusters
at the distances of the Magellanic Clouds causes significant systematic
photometric uncertainties that are largely absent in HST photometry.   

The one cluster without clear evidence for an eMSTO is Hodge 6, for which the
empirical pseudo-age distribution is only marginally wider than that of its SSP
simulation. This is likely due at least in part to it being the oldest cluster in
our sample (age = 2.25 Gyr). Since the width of pseudo-age distributions  
of simulated SSPs in Myr scales approximately with the logarithm of the
clusters' age \citep{goud+11a,kell+11}, the ability to detect a given age
spread is age dependent, becoming harder for older clusters.  

\subsection{Relation to Star Formation Histories} 
\label{sub:SFHs}

We emphasize that the observed pseudo-age distributions shown in the third row
of panels in Figs.\ \ref{f:cmdplot1}\,--\,\ref{f:cmdplot3} do \emph{not} reflect
the clusters' star formation histories (SFHs) in case of clusters with
non-negligible levels of initial mass segregation. This is due to the strong
``impulsive'' loss of stars taking place after the massive stars in the inner
regions of mass-segregated clusters reach their end of life, which causes the
cluster to expand beyond its tidal radius, thereby stripping its outer layers 
\citep[e.g.,][]{vesp+09}. In the context of the ``in situ'' scenario,
this loss would mainly occur for the first generation of stars, as the
second generation is formed in the innermost regions of the cluster after
the impulsive loss of first-generation stars has finished \citep{derc+08}.  

To estimate SFHs from the observed pseudo-age distributions of these star
clusters in the context of the ``in situ'' scenario, one needs to consider
\emph{(i)} the evolution of the number of first-- and second-generation stars
from the clusters' birth to the current epoch, and \emph{(ii)} the age
resolution element of the pseudo-age distributions, i.e., the shape of the
function that describes a SSP in the pseudo-age distributions. For the latter,
we use a gaussian with a FWHM equal to that of the pseudo-age distribution of
the SSP simulation of the cluster in question (i.e., the red curves in the third
column of panels in Figs.\ \ref{f:cmdplot1}\,--\,\ref{f:cmdplot3}). 

To evaluate consideration \emph{(i)} above for the case of initially
mass-segregated clusters, we again adopt the results of the simulation called
SG-R1 in \citet[][see their Fig.\ 15]{derc+08}. 
However, rather than using a given fixed level of initial mass segregation for
every cluster, we consider it likely that this level varied among clusters. This
implies that the (time-dependent) number ratio of initial-to-current first-generation stars
(defined here as ${\it NR}_{\rm FG} (t) \equiv N_{\rm FG}^{\rm init} / N_{\rm FG} (t)$) also varies
among clusters.  
To estimate a \emph{plausible} value of ${\it NR}_{\rm FG} (t)$ for each
cluster, we use results of the study by \citet{mack+08b} who showed that the
maximum core radius seen among a large sample of Magellanic Cloud star clusters
increases approximately linearly with log\,(age) up to an age of about 1.5 Gyr,
namely from $\simeq$\,2.0 pc at $\simeq$\,10 Myr to  $\simeq$\,5.5 pc at
$\simeq$\,1.5 Gyr. In contrast, the \emph{minimum} core radius is about 1.5 pc
throughout the age range 10 Myr\,--\,2 Gyr. N-body modeling by \citet{mack+08b}
showed that this behavior is consistent with the adiabatic expansion of the
cluster core in clusters with varying levels of initial mass segregation, in the
sense that clusters with the highest level of initial mass segregation
experience the strongest core expansion. 

Another result of the simulations by \citet{mack+08b} that is relevant to the
current discussion is the impact of the retention of stellar black holes (BHs)
to the evolution of the clusters' core radii. As shown in their Figs.\ 5, 15,
and 21, simulated clusters that are able to retain the BHs formed earlier by
stellar evolution of the massive stars experience a continuation of core
expansion at ages $\ga$\,1 Gyr, due to superelastic collisions between BH
binaries and other BHs in the central regions. In contrast, clusters that do
not retain stellar BHs start a slow core contraction process at an age of
$\sim$\,1 Gyr due to two-body relaxation. While the currently available data do
not allow direct constraints on the BH retention fraction of the clusters in
our sample, the results of the simulations by \citet{mack+08b} do imply that
the large core radii of clusters with ages in the approximate range of 1\,--\,2 
Gyr and core radii $\rc \ga 5.5$ pc do not necessarily indicate
extraordinarily high levels of initial mass segregation, even though their
levels of initial mass segregation are likely still higher than for clusters in
that age range that have $\rc \la 3.5$ pc. 
A more complex degeneracy is present for clusters in the age range of
$\sim$\,2\,--\,3 Gyr with $3.5 \la \rc/{\rm pc} \la 5.5$. Such clusters can
be produced by simulations of clusters \emph{without} initial mass segregation
that \emph{do} retain their BHs just as well as by simulations \emph{with}
significant levels of initial mass segregation that do \emph{not} retain their
BHs (see Fig.\ 5 in \citealt{mack+08b}).  

With this in mind, we tentatively assign values of ${\it NR}_{\rm FG} (t)$ 
to each cluster in the following way\footnote{We emphasize that the application of this
  procedure is formally specific to star clusters in the Magellanic Clouds. It
  may or may not be applicable to other environments.}. For clusters with $\rc
\leq 5.5$ pc and an age in the range 1\,--\,2 Gyr, we assume that the (current)
size of the core radius reflects the level of initial mass segregation of the
cluster and we set the ``plausible'' value of ${\it NR}_{\rm FG} (t)$ as follows:
\begin{equation}
{\it NR}_{\rm FG}^{p} (t) \equiv {\rm max}\left(1, {\it NR}_{\rm FG}^{\rm seg} (t)
  \; \times \left( \frac{\rc-1.5}{5.5-1.5} \right) \right)\mbox{.} 
\label{eq:NR_FG}
\end{equation}
where ${\it NR}_{\rm FG}^{\rm seg} (t)$ is the number ratio of initial-to-current
first-generation stars calculated for the case of the SG-R1 model of
\citet{derc+08}. 
For clusters in our sample with $\rc > 5.5$ pc and ages $>$\,1.5 Gyr, 
we hypothesize that the core radius may have increased due in part to dynamical
effects related to the presence of stellar-mass black holes in the central
regions. This makes it hard to relate the current core radius to a 
particular level of initial mass segregation, even though this level is most
likely still substantial. Hence we simply estimate 
${\it NR}_{\rm FG}^{p} (t) \equiv {\it NR}_{\rm FG}^{\rm seg} (t) \times 2/3$
for such clusters. 
Recognizing that the values of ${\it NR}_{\rm FG}^{p} (t)$ for these clusters are
inherently more uncertain than for those with $\rc \leq 5.5$ pc, we plot the
resulting SFHs with cyan lines in Figs.~\ref{f:cmdplot1}\,--\,\ref{f:cmdplot3}
(right-hand panels). 
Finally, SFHs of clusters with ages $\geq 2$ Gyr and $3.5 \leq \rc/{\rm pc} <
5.5$ are assigned magenta lines in Figs.~\ref{f:cmdplot1}\,--\,\ref{f:cmdplot3}. 

The SFH of a given cluster is then estimated by applying the inverse evolution
of the number of first-generation stars (i.e., the multiplicative factor
${\it NR}_{\rm FG}^{p} (t)$) to the ``oldest'' resolution element of the pseudo-age
distribution, i.e., a resolution element whose wing on the right-hand (``old'')
side lines up with the ``oldest'' non-zero part of the cluster's pseudo-age
distribution.  In contrast, the inverse evolution of the number of
second-generation stars is applied to the full pseudo-age distribution. The
resulting SFHs are shown in the right-hand panels of Figs.\
\ref{f:cmdplot1}\,--\,\ref{f:cmdplot3}.  
For the estimated levels of initial mass segregation of the clusters in our
sample, our results indicate that the SFR of the first generation dominated
that of the second generation by factors between about 3 and 10.

\subsection{Relations Between MSTO Width and Early Dynamical Properties} 
\label{sub:correl}

To quantify the differences between the pseudo-age distributions of the cluster
data and those of their SSP simulations in terms of {\it intrinsic\/}
MSTO widths of the clusters, we measure the widths of the two sets of
distributions at 20\% and 50\% of their maximum values (hereafter
called W20 and FWHM, respectively), using quadratic interpolation. The
intrinsic pseudo-age ranges of the clusters are then estimated by
subtracting the simulation widths in quadrature: 
\begin{eqnarray}
{\it W20}_{\rm MSTO} & = & ({\it W20}_{\rm obs}^2 - {\it W20}_{\rm SSP}^2)^{1/2} \nonumber \\
{\it FWHM}_{\rm MSTO} & = & ({\it FWHM}_{\rm obs}^2 - {\it FWHM}_{\rm
  SSP}^2)^{1/2} 
\label{eq:widths}
\end{eqnarray}
where the ``obs'' subscript indicates measurements on the observed CMD and the
``SSP'' subscript indicates measurements on the simulated CMD for a SSP. 
Given the insignificant difference between the width of the MSTO of Hodge 6 and
that of its SSP simulations, we designate its resulting values for ${\it
  FWHM}_{\rm MSTO}$ and ${\it W20}_{\rm MSTO}$ as upper limits. The same is done
for the lower-mass LMC clusters NGC 1795 and IC 2146 (with ages of 1.4 and 1.9 Gyr,
respectively) for which \citet{corr+14} finds their MSTO widths to be consistent
with those of their SSP simulations to within the uncertainties \citep[see
also][]{milo+09}, even though two other intermediate-age LMC clusters with
  similarly low masses (but different structural parameters) \emph{do} exhibit
  eMSTOs.   

\subsubsection{A Correlation Between MSTO Width and Early Cluster Mass} 

\begin{figure*}[tbh]
%\centerline{\includegraphics[width=14cm]{f5.ps}}
\centerline{\includegraphics[width=11cm]{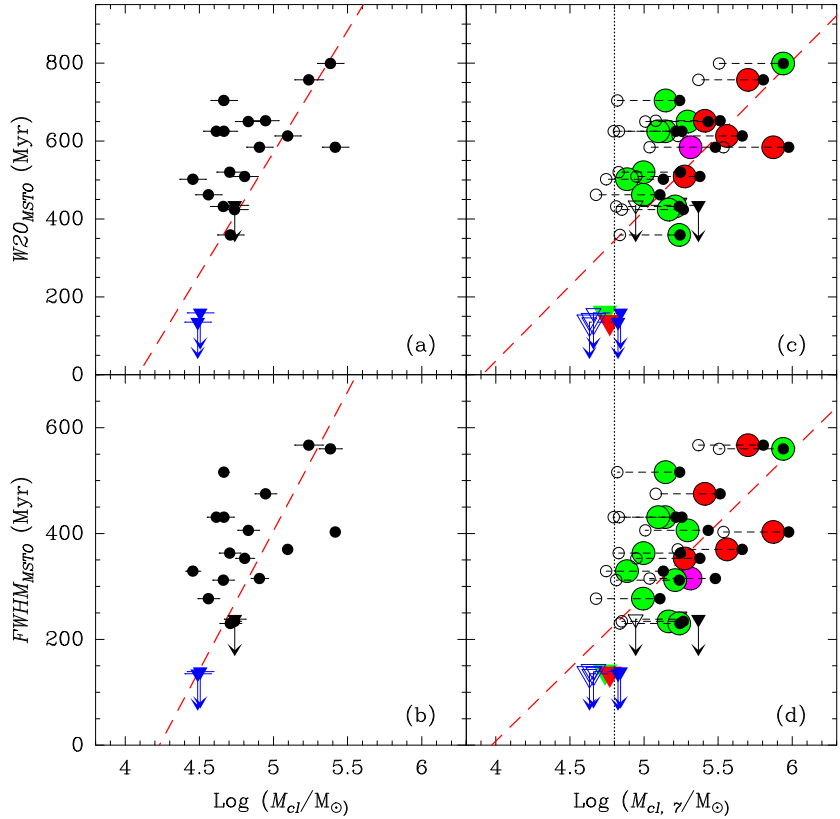}}
\caption{\emph{Panel (a)}: {\it W20}$_{\it MSTO}$ versus $\cal{M}_{\rm cl}$, the
  current cluster mass. Upper limits to {\it W20}$_{\it
    MSTO}$ are indicated by triangles and downward arrows. Black symbols
  represent clusters in our sample, while dark blue symbols represent NGC 1795
  and IC 2146, two low-mass clusters from the \citet{milo+09} sample. The red
  dashed line represents a linear regression fit to the data using the
  Buckley-James method which takes upper limits into account
  \citep[see][]{feinel85}. \emph{Panel (b)}: Similar to panel (a), but now {\it
    FWHM}$_{\it MSTO}$ versus $\cal{M}_{\rm cl}$. 
  \emph{Panel (c)}: Similar to panel (a), but now {\it W20}$_{\it MSTO}$ versus
  $\cal{M}_{\rm cl,\,7}$, the cluster mass calculated for an
  age of 10 Myr (see \S\,\ref{s:dynevol}). Small open and
  small filled symbols indicate values of $\cal{M}_{\rm cl,\,7}$ for models without
  and with initial mass segregation, respectively. Large symbols in green, red,
  or magenta indicate educated guesses for the actual values of $\cal{M}_{\rm
    cl,\,7}$ for the clusters (see discussion in \S\,\ref{sub:correl} for the
  meaning of the colors). The dotted line indicates log($\cal{M}_{\rm cl,\,7}$) = 
  4.8, which seems to represent the approximate early cluster mass above which
  clusters seem to be able to host an eMSTO. The red dashed line represents a linear
  regression fit to the large green, red, and magenta symbols.  
  \emph{Panel (d)}: Similar to panel (c), but now {\it FWHM}$_{\it MSTO}$
  versus $\cal{M}_{\rm cl,\,7}$. See discussion in \S\,\ref{sub:correl}.1.
}
\label{f:massplot}
\end{figure*}

We plot ${\it W20}_{\rm MSTO}$ and ${\it FWHM}_{\rm MSTO}$ versus $\cal{M}_{\rm
    cl}$ at the current age in Figures~\ref{f:massplot}a and \ref{f:massplot}b,
  respectively. Note that the width of the MSTO region seems to correlate with
  the cluster's current mass. To quantify this impression, we perform
  statistical tests for the probability of a correlation in the presence of
  upper limits, namely the Cox Proportional Hazard Model test and the generalized
  Kendall's $\tau$ test (see \citealt{feinel85}). The results are listed in
  Table~\ref{t:correl}. The probabilities of the absence of a
  correlation are small: $p < 1.3$\%. However, we
  remind the reader that the clusters plotted here have different ages and
  radii, and hence likely underwent different amounts of mass loss since their
  births. This complicates a direct interpretation of this correlation in terms
  of constraining formation scenarios. To estimate the nature of this relation
  at a cluster age of 10 Myr, we therefore also plot ${\it W20}_{\rm
    MSTO}$ and ${\it FWHM}_{\rm MSTO}$ against $\cal{M}_{\rm cl,\,7}$ in
  Figs.~\ref{f:massplot}c and \ref{f:massplot}d, respectively.    
In view of the uncertainty of assigning initial levels of mass segregation to
individual clusters, we consider a {\it range\/} of possible $\cal{M}_{\rm
  cl,\,7}$ values for each cluster, shown by dashed horizontal lines in
Figs.~\ref{f:massplot}c and \ref{f:massplot}d. The minimum and
maximum values of $\cal{M}_{\rm cl,\,7}$ for each cluster are the values
resulting from the calculations without and with initial mass
segregation, respectively. These values will be called $\cal{M}_{\rm
  cl,\,7}^{\rm noseg}$ and $\cal{M}_{\rm cl,\,7}^{\rm seg}$ hereinafter.  

As seen in Figs.~\ref{f:massplot}c and \ref{f:massplot}d, the range
of possible $\cal{M}_{\rm cl,\,7}$ values for a given cluster can be
significant. This is especially so for the older clusters in our sample, owing
to the longer span of time during which the cluster has experienced mass
loss. To estimate a plausible value of $\cal{M}_{\rm esc,\,7}$ for each
cluster, we follow the arguments based on 
the results of the \citet{mack+08b} study described in the previous section. For
clusters with $\rc \leq 5.5$ pc and an age $\leq$~2 Gyr, we thus set the
``plausible'' value of $\cal{M}_{\rm cl,\,7}$  as follows:
\begin{equation}
{\cal{M}}_{\rm esc,\,7}^{p} \equiv {\cal{M}}_{\rm cl,\,7}^{\rm noseg} +
({\cal{M}}_{\rm cl,\,7}^{\rm seg} - {\cal{M}}_{\rm cl,\,7}^{\rm noseg})\; \times 
\left( \frac{r_{\rm c}-1.5}{5.5-1.5} \right)\mbox{.} 
\label{eq:plausible}
\end{equation}
These values of ${\cal{M}}_{\rm cl,\,7}^{p}$ are shown by large green
  symbols in Figs.~\ref{f:massplot}c and \ref{f:massplot}d. For clusters in our
  sample with $\rc > 5.5$ pc and ages $>$\,1.5 Gyr, we estimate ${\cal{M}}_{\rm
    cl,\,7}^{p} \equiv ({\cal{M}}_{\rm cl,\,7}^{\rm noseg} + {\cal{M}}_{\rm
    cl,\,7}^{\rm seg} \times 2)/3$ and we plot them with large red symbols in
  Figs.~\ref{f:massplot}c and   \ref{f:massplot}d. Finally, ${\cal{M}}_{\rm
    cl,\,7}^{p}$ values of clusters with ages $\geq 2$ Gyr and $3.5 \leq
  \rc/{\rm pc} < 5.5$ are assigned large magenta symbols in
  Figs.~\ref{f:massplot}c and \ref{f:massplot}d. We also included the results
  for the two low-mass LMC clusters NGC 1795 and IC 2146 studied by \citet{corr+14}.  

The distribution of the ``large'' symbols in Figs.~\ref{f:massplot}c and
\ref{f:massplot}d again reveals a correlation between the width of the MSTO
region and the early cluster mass. 
Results of correlation tests are listed in Table~\ref{t:correl}. The 
correlations between MSTO width and early cluster mass are significant,
with $p$ values that are $\simeq$\,50\% smaller than those for the relations
of the FWHM widths with the \emph{current} cluster masses.

Comparing the early masses of the clusters that exhibit eMSTOs with
those that do not, our results indicate that the ``critical mass'' needed for
the creation of an eMSTO is around log(${\cal{M}}_{\rm cl,\,7}) \approx 4.8$,
indicated by a dotted line in Figs.~\ref{f:massplot}c and \ref{f:massplot}d. 
This contrasts with the predictions of the ``in situ'' star formation model of
\citet{conspe11} in terms of the minimum mass required for clusters to be able 
to accrete pristine gas from the surrounding ISM in the LMC environment, which
they estimated to be $\approx 10^4\; M_{\odot}$. In other words, our results
indicate that this minimum mass may be a factor $\approx 6-8$ higher than that
estimated by \citet{conspe11}\footnote{This factor would be $\approx 4-5$ when
  using the Kroupa IMF.}. 
  However, the data in Fig.~\ref{f:massplot} do not reveal an \emph{obvious}
  minimum mass threshold in this context, and some young star clusters in the
  LMC without signs of an eMSTO are more massive than this (e.g., NGC 1856 and
  NGC 1866: ${\cal{M}}_{\rm cl} \sim 10^5 \; M_{\odot}$,
  \citealt{bassil13}). This may indicate that other properties (in addition
  to the early cluster mass) are relevant in terms of the ability of star
  clusters to accrete gas from their surroundings (e.g., the cluster's
  velocity relative to that of the surrounding ISM, cf.\ \citealt{conspe11},
  and the actual local distribution of ISM at the time).

\subsubsection{A Correlation Between MSTO Width and Early Escape Velocity} 

\begin{figure*}[tbh]
%\centerline{\includegraphics[width=14cm]{f6.ps}}
\centerline{\includegraphics[width=11cm]{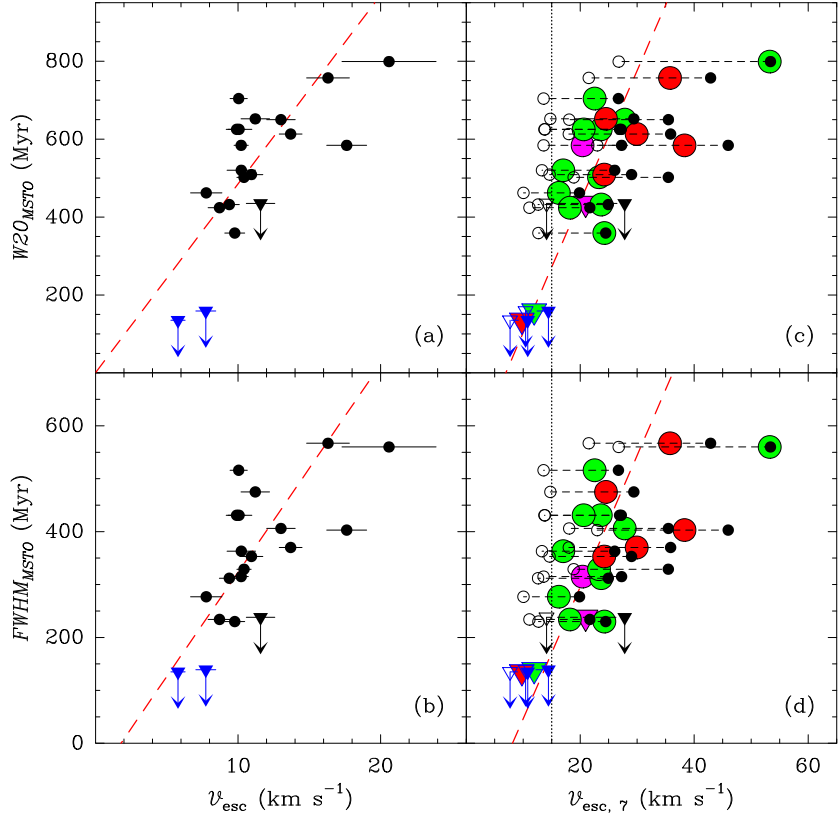}}
\caption{\emph{Panel (a)}: {\it W20}$_{\it MSTO}$ versus $v_{\rm esc}$, the
  current escape velocity at the core radius. Symbols are the same as for
  Figure~\ref{f:massplot}a. The red dashed line represents a formal linear
  inverse-variance weighted fit to the data. \emph{Panel (b)}: Similar to panel
  (a), but now {\it FWHM}$_{\it MSTO}$ versus $v_{\rm esc}$. 
  \emph{Panel (c)}: Similar to panel (a), but now {\it W20}$_{\it MSTO}$ versus
  $v_{\rm esc,\,7}$, the escape velocity at the core radius calculated for an
  age of 10 Myr (see \S\S\,\ref{s:dynevol} and \ref{s:vesc}). Symbols are the
  same as for Figure~\ref{f:massplot}c. The dotted line indicates $v_{\rm
    esc,\,7}$ = 15 \kms. The red dashed line represents a formal linear fit to 
  the large green, red, and magenta symbols. 
  \emph{Panel (d)}: Similar to panel (c), but now {\it FWHM}$_{\it MSTO}$
  versus $v_{\rm esc,\,7}$. See discussion in \S\,\ref{sub:correl}.2.
}
\label{f:Vescplot}
\end{figure*}

We plot ${\it W20}_{\rm MSTO}$ and ${\it FWHM}_{\rm MSTO}$ versus $v_{\rm
    esc}$ at the current age in Figures~\ref{f:Vescplot}a and \ref{f:Vescplot}b,
  respectively. Similar to the correlation with cluster mass shown in
  Figures~\ref{f:massplot}, the width of the MSTO region correlates with
  the cluster's current escape velocity. In this case, the probabilities of the
  absence of a correlation are $p \la 0.3$\% (see Table~\ref{t:correl}), which
  is significantly lower than for the relation between MSTO width versus cluster
  mass.  

  To estimate the nature of the relation of the MSTO width with escape velocity 
  at a cluster age of 10 Myr, we plot ${\it W20}_{\rm
    MSTO}$ and ${\it FWHM}_{\rm MSTO}$ against $v_{\rm esc,\,7}$ in
  Figs.~\ref{f:Vescplot}c and \ref{f:Vescplot}d, respectively.    
``Plausible'' values of $v_{\rm esc,\,7}$ were determined using the
assignments of levels of initial mass segregation and its
associated scaling relations described above in \S\,\ref{sub:correl}.1. 

% Place Table 4 here.
\addtocounter{table}{1}
\begin{table}[tbh]
\footnotesize
\caption[]{Results of correlation tests.}
\label{t:correl}
\begin{center}
\begin{tabular}{@{}lcccc@{}} \tableline \tableline
\multicolumn{3}{c}{~~} \\ [-2ex]  
\multicolumn{1}{c}{Relation} & $p_{\rm cox}$ & $\tau$ & $Z$ & $p_{\tau}$ \\
\multicolumn{1}{c}{(1)} & (2) & (3) & (4) & (5)\\ [0.5ex] \tableline 
\multicolumn{3}{c}{~~} \\ [-1.5ex] 
    {\it W20}$_{\it MSTO}$ vs.\ log(${\cal{M}}_{\rm cl}$) & 0.0006 & 0.8737 & 2.712 & 0.0067 \\
   {\it FWHM}$_{\it MSTO}$ vs.\ log(${\cal{M}}_{\rm cl}$) & 0.0006 & 0.8000 & 2.479 & 0.0132 \\
 {\it W20}$_{\it MSTO}$ vs.\ log(${\cal{M}}_{\rm cl,\,7}^{p}$) & 0.0013 & 0.8947 & 2.778 & 0.0055 \\
{\it FWHM}$_{\it MSTO}$ vs.\ log(${\cal{M}}_{\rm cl,\,7}^{p}$) & 0.0012 & 0.8211 & 2.544 & 0.0110 \\
        {\it W20}$_{\it MSTO}$ vs.\ $v_{\rm esc}$ & 0.0002 & 0.9263 & 2.877 & 0.0040 \\
       {\it FWHM}$_{\it MSTO}$ vs.\ $v_{\rm esc}$ & 0.0002 & 0.9684 & 3.001 & 0.0027 \\
 {\it W20}$_{\it MSTO}$ vs.\ $v_{\rm esc,\,7}^{p}$ & 0.0001 & 1.0421 & 3.235 & 0.0012 \\
{\it FWHM}$_{\it MSTO}$ vs.\ $v_{\rm esc,\,7}^{p}$ & 0.0001 & 0.9789 & 3.035 & 0.0024 \\ [0.5ex] \tableline
\multicolumn{3}{c}{~~} \\ [-1.2ex]
\end{tabular}                   
\tablecomments{Column (1): relation being tested. Column (2): probability of
  an absence of a correlation according to the Cox Proportional Hazard
  Model test. Column (3): value of generalized Kendall's correlation
  coefficient. Column (4): $Z$-value of 
  generalized Kendall's correlation test. Column (5): probability of
  an absence of a correlation according to the generalized Kendall's correlation
  test. See discussion in \S\,\ref{sub:correl}.
}
\end{center}
%\parbox{8.3cm}{
%{\sl Notes}.~~Column (1): NGC number of galaxy. 
%Column (2): absolute $V$ magnitude of galaxy. Column (3:) ${\it col}_0$ of galaxy. 
%Column (4): ${\it col}_0$ of red GCs. Column (5): color gradient $G_{\it col}$ of red GCs.   
%Column (6): mean rms error of fit of Eq.\ (1) to colors of red GCs
%(in mag).}
\end{table}

The distribution of the ``large'' symbols in Figures~\ref{f:Vescplot}c and
\ref{f:Vescplot}d again shows that {\it FWHM}$_{\it MSTO}$ (or {\it W20}$_{\it
MSTO}$) is correlated with $v_{\rm esc,\,7}^{p}$ in that clusters with larger
early escape velocities have wider MSTO regions. 
A glance at Table~\ref{t:correl} shows that these correlations are highly
  significant, with $p$-values that are about half of those for the relations of
  the MSTO widths with the \emph{current} escape velocities.  
  The $p$-values for the relation between MSTO width and escape velocity
  are also significantly lower than those for the relation between MSTO width
  and cluster mass, suggesting a more causal correlation for the former. 
Finally, Figs.~\ref{f:Vescplot}c and \ref{f:Vescplot}d suggest that
  eMSTOs occur only in clusters with early escape velocities $v_{\rm esc,\,7}
  \ga$~12\,--\,15 \kms.

\section{Discussion}
\label{s:disc}

We review how our results compare with recent predictions of the ``stellar
rotation'' and the ``in situ star formation'' scenarios below. 
  We also compare our results with relevant findings in the recent
  literature, and comment on the feasibility of other scenarios to
  explain the eMSTO phenomenon in the light of our results. Finally, we
  discuss our results in the context of the currently available data on
  light-element abundance variations within the clusters in our sample. 

\subsection{Comparison with Stellar Rotation Scenario}
\label{s:inter_rot}

\subsubsection{MSTO Widths}

\begin{figure}[tbh]
%\centerline{\includegraphics[width=8cm]{rotationplot.ps}}
\centerline{\includegraphics[width=7.cm]{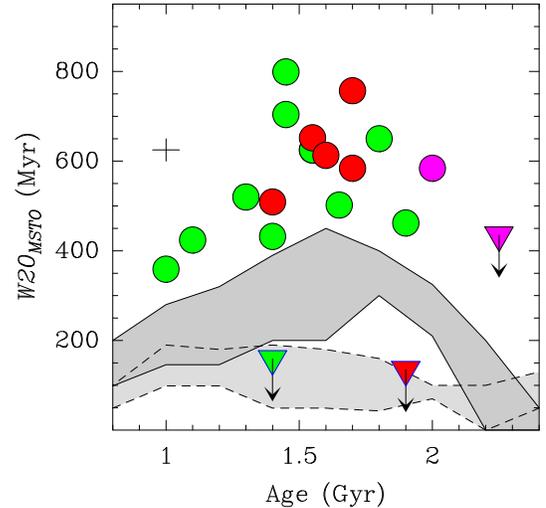}}
\caption{{\it W20}$_{\it MSTO}$ versus cluster age. Symbols are the
  same as the ``large'' symbols in Fig.~\ref{f:Vescplot}. The grey
  areas delimited by solid and dashed lines represent the ranges of MSTO width
  as function of age due to stellar rotation effects found by \citet{yang+13}
  for mixing efficiencies $f_c$ = 0.03 and 0.20, respectively. See discussion
  in \S\,\ref{s:inter_rot}. 
}
\label{f:W20compare}
\end{figure}

To compare the MSTO widths with predictions of the stellar rotation scenario,
we use results from the recent study of \citet{yang+13} 
who calculated evolutionary tracks of non-rotating and rotating stars for three 
different initial stellar rotation periods (approximately 0.2, 0.3, and 0.4
times the Keplerian rotation rate of ZAMS stars), and for two different mixing
efficiencies (``normal'', $f_c$ = 0.03, and ``enhanced'', $f_c$ = 0.20). From
the isochrones built from these tracks, they calculated the widths of the MSTO
region caused by stellar rotation as a function of cluster age and translated
them to age spreads (in Myr). In the context of the pseudo-age distributions
derived for our clusters in Section~\ref{sub:MCsim}, the age spreads due to
rotation calculated by  \citet{yang+13} are equivalent to the \emph{full}
widths of the age distribution (W.\ Yang, 2014, private communication). Hence
we compare their age spreads with our W20 values. 
Using the results shown in Fig.\ 8 of \citet{yang+13}, 
we assemble the ranges encompassed by their age spreads as a function of age
for the two different mixing efficiencies\footnote{Specifically, we assemble the
  minima and maxima of the equivalent age spreads plotted by them for the 3
  rotation periods 0.37, 0.49, and 0.73 days.}. These ranges are shown as grey
regions delimited by solid and dashed curves in
Figure~\ref{f:W20compare}, which shows {\it W20}$_{\it MSTO}$ as function of age for
the clusters in our sample.  

Figure~\ref{f:W20compare} reveals some interesting results. First of all, many
clusters in our sample feature MSTO widths that are significantly larger
than stellar rotation seems to be able to produce at their age according to the
\citet{yang+13} study. This result, along with the finding that {\it W20}$_{\it MSTO}$
correlates with $v_{\rm esc,\,7}$, suggests strongly that \emph{the eMSTO
  phenomenon is at least partly due to ``true'' age effects}. Second, stellar
rotation at ``normal'' mixing efficiency seems to be able to produce age spreads
that generally follow the lower envelope of the measured {\it W20}$_{\it
  MSTO}$ values of the star clusters in our sample as a function of their age.  
This by itself could indicate that the MSTO broadening of these
clusters could also be due in part to stellar rotation, 
for example if some clusters might host significant numbers of stars with
  rotation rates $>$\,0.4 times the Keplerian rate.  
However, the two clusters with ages in the 1\,--\,2 Gyr range that do not show
any measurable amount of MSTO broadening (NGC~1795 and IC~2146) are inconsistent
with this view, \emph{unless} stellar rotation in such clusters (with low 
values of $v_{\rm esc,\,7}$) either occurs at much lower rotation velocities or
at significantly higher mixing efficiency than in clusters with higher values
of $v_{\rm esc,\,7}$.    

In this context we note two reasons suggesting that stellar
rotation rates should \emph{not} be significantly different in those two
low-mass clusters relative to the other clusters in our sample: 
\emph{(i)} the absolute difference in $v_{\rm esc,\,7}$ between those two
clusters and the lowest-mass clusters in our sample that \emph{do} feature
eMSTOs is not huge ($\approx$\,9\,--\,12 vs.\ 16\,--\,20 \kms). This
  difference is even smaller when considering their \emph{current} masses or
  escape velocities (see Figs.\ \ref{f:massplot} and
  \ref{f:Vescplot}). Hence, the onset of the widening of the MSTO seems more
likely to be caused by a ``minimum'' threshold escape velocity (at early
times) than by pure relative depths of the clusters' potential wells (see also
\citealt{corr+14} and \S\,\ref{s:inter_sf} 
below).   
\emph{(ii)} In a recent study of the Galactic open cluster Trumpler 20, 
the only star cluster in the age range of 1\,--\,2 Gyr for which stellar
rotation velocities have been measured to date using high-resolution
spectroscopy, \citet{plat+12} found an approximately flat distribution of
rotation velocities of MSTO stars in the range $180 < V \sin i < 0$ \kms. This
implies a range of rotation rates very similar to that considered by the
\citet{yang+13} models, even though this is a loose cluster with very low escape
velocity. Furthermore, \citet{plat+12} found that the 50\% fastest rotators in
Trumpler 20 are actually marginally \emph{blueshifted} on the CMD with respect
to the slow rotators ($\delta(V-I) = -0.01$, see \citealt{plat+12}).  
These findings are inconsistent with the predictions of \citet{yang+13} for
``normal'' mixing efficiency, but they are marginally consistent with their
predictions for high mixing efficiency (see Fig.~\ref{f:W20compare}). While it
is not clear to us why the efficiency of rotational mixing in stars would be
higher in clusters with lower potential well depths, this may be an avenue for
future research. 
  We also recognize that the study of the creation of theoretical
  stellar tracks and isochrones for rotating stars at various stages of
  stellar evolution, rotation rates, and ages is still in relatively early
  stages, and that our comparison with model predictions such as those of
  \citet{yang+13} implicitly involves adopting the assumptions made by those
  models. Furthermore, no stellar rotation measurements have yet been 
  undertaken in intermediate-age star clusters in the Magellanic
  Clouds. Hence, future findings in this context might affect our conclusions 
  on the nature of eMSTOs. 
However, for now, the observations of \citet{plat+12} are most consistent
with the predictions of \citet{gira+11}, i.e., that stellar models with
rotation produce a marginal blueshift in the MSTO of star clusters with ages in
the range 1\,--\,2 Gyr rather than a reddening as predicted by \citet{basdem09}
and \citet{yang+13}.  

\subsubsection{Red Clump Morphologies}

Focusing on the RC feature in the CMDs of the star clusters in our sample as a
function of their (average) age, one can identify certain trends that are
relevant to the nature of eMSTOs. Firstly, the RC feature can often be seen to
extend to fainter magnitudes than the RC feature of the clusters' respective SSP 
simulations. This is especially the case for the relatively  
massive clusters NGC 411, NGC 419, NGC 1852, and NGC 2203; hints of this effect
also appear in NGC 2154, and NGC 2173. This ``composite red clump''
feature was already reported before in NGC 411, NGC 419, NGC 1751, NGC 1783, and
NGC 1846 \citep{gira+09,gira+13,rube+10,rube+13}, and is thought to be due to
the cluster hosting stars massive enough to avoid electron degeneracy settling
in their H-exhausted cores when He ignites. The main part of the RC
consists of less massive stars which did pass through electron
degeneracy prior to He ignition \citep{gira+09}. This causes the
brightness of RC stars to increase relatively rapidly with decreasing stellar
mass in the narrow age range of $\simeq$\,1.00\,--\,1.35 Gyr, after which that
increase slows down significantly due to the fact that \emph{all} RC stars
experienced electron degeneracy prior to He ignition. 
Interestingly, the composite RCs are seen in \emph{all} clusters in our sample
for which the pseudo-age distributions indicate 
the presence of a non-negligible number of stars in that age range,
even though their best-fit age is always older than 1.35
Gyr\footnote{Another cluster with an extended RC is Hodge 2 with a
  best-fit age of 1.3 Gyr. At this age, the RC is naturally extended
  \citep[see][]{gira+09} rather than ``composite''. However, the left 
  panel of Fig.~\ref{f:cmdplot2} for Hodge 2 does show that its RC feature
  extends to fainter magnitudes than that of its best-fit isochrone. This is
  consistent with hosting stellar ages younger than 1.3 Gyr, as indicated by
  its pseudo-age distribution.}. 

\begin{figure*}[tbh]
%\centerline{\includegraphics[width=14cm]{f8.ps}}
\centerline{\includegraphics[width=12.5cm]{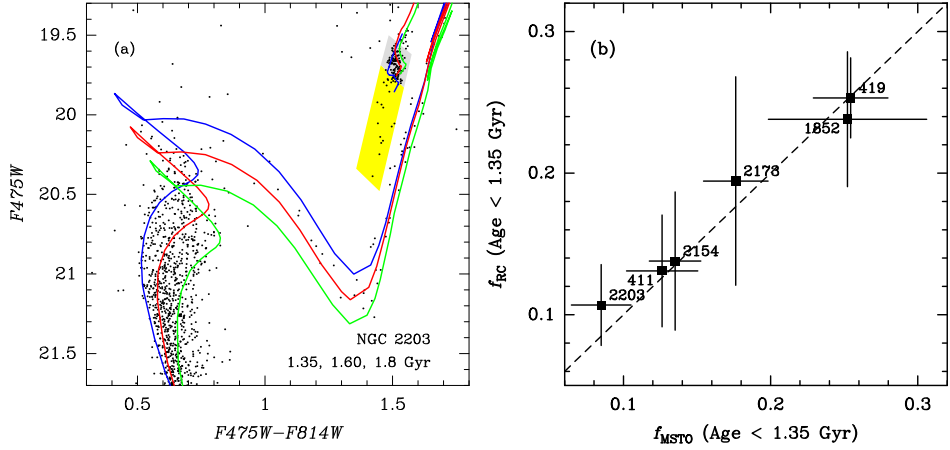}}
\caption{\emph{Panel (a)}: CMD of stars within the core radius of NGC 2203,
  zooming in on the MSTO and RC regions. The blue, red, and green lines depict
  \citet{mari+08} isochrones with ages 1.35, 1.60, and 1.85 Gyr,
  respectively. The yellow parallelogram depicts the faint ``secondary RC''
  region described in \S\,\ref{s:inter_rot}, while the grey parallelogram
  depicts the part of the ``full RC'' region that is not part of the ``secondary
  RC'' region.   
  \emph{Panel (b)}: the fraction of RC stars in the secondary RC plotted
  against the fraction of MSTO stars with ages $\leq$\,1.35 Gyr for
  clusters whose pseudo-age distribution shows a significant number of stars
  with ages $\leq$\,1.35 Gyr. The NGC numbers of the clusters are indicated next
  to their data points. The dashed line represents a 1:1 relation. See
  discussion in \S\,\ref{s:inter_rot}.
}
\label{f:RCplot}
\end{figure*}

This has an impact to the feasibility of the ``stellar rotation'' 
scenario in causing the eMSTOs for these clusters. In this scenario, stars with
high rotation velocities have larger core masses at the end of the MS era than
do non-rotating stars \citep[e.g.,][]{maemey00,egge+10,yang+13}. 
In this respect, fast rotators could in principle present the modest
increase in the core mass necessary to ignite Helium before the
settling of electron degeneracy, and hence cause the faint
extension of the RC as well. However, the 
(small) fraction of RC stars in its faint extension scales with the
fraction of MSTO stars at the \emph{youngest} ages in these clusters, i.e., at
ages in the 1.00\,--\,1.35 Gyr range, which are at the \emph{blue} and bright end
of the MSTO. 
This is illustrated in Figure~\ref{f:RCplot}. Figure~\ref{f:RCplot}a shows
  the CMD of NGC 2203 (as an example) along with three isochrones for ages 1.35,
  1.60, and 1.85 Gyr. These isochrones coincide approximately with the upper
  left edge, the mean location, and the lower right edge of its eMSTO feature,
  respectively. The RC and AGB parts of the same isochrones are also shown on
  top of the RC of NGC 2203. The faint ``secondary RC'' feature
  \citep[cf.\ ][]{gira+09}, which is shown as a yellow parallelogram, is then
  defined as the area in the CMD ``below'' the horizontal branch (HB) of the
  1.35 Gyr isochrone; the tilt of the short side of the parallelogram equals
  that of the HB of the 1.35 Gyr isochrone. The ``full RC'' area is then
  approximated by extending the ``secondary RC'' area toward brighter magnitudes
  so as to also encompass the full HB of the ``oldest'' isochrone, allowing for 
  suitable photometric errors. (This extension is shown as a light grey
  parallelogram in Fig.~\ref{f:RCplot}a.) We then evaluate the fraction of RC
  stars in the secondary RC, defined as $f_{\rm RC} (< 1.35\;{\rm Gyr})
  \equiv N\mbox{(secondary RC)}/N\mbox{(full RC)}$. This fraction is plotted versus
  $f_{\rm MSTO} (< 1.35\;{\rm Gyr})$, the fraction of stars in the pseudo-age
  distributions at ages $\leq$~1.35 Gyr, in Figure~\ref{f:RCplot}b for all
  clusters whose pseudo-age distribution in
  Figs.~\ref{f:cmdplot1}\,--\,\ref{f:cmdplot3} indicates the presence of a
  significant number of stars with ages $\leq$~1.35 Gyr even though their
  average age is older. Note that even though the Poisson uncertainties are
  significant, the data indicate an approximate 1:1 relation between the
  fraction of RC stars in the faint extension and the fraction of MSTO stars in
  the part on the ``upper left'' side of the 1.35 Gyr isochrone. 
Note that the sense of this relation is consistent with that predicted in case
the width of eMSTOs reflects a range in stellar ages, while it is contrary to  
the predictions of \citet{basdem09} and \citet{yang+13} which 
were that high stellar rotation velocities cause stars to populate the
\emph{lower right} end of the MSTO at the ages of these clusters. This
suggests that eMSTOs are indeed due mainly to a range of stellar ages
rather than a range of stellar rotation velocities among MSTO stars. 
  This result would benefit from confirmation by future isochrones of  
  rotating stars that include the stages of stellar evolution past the Helium
  flash on the RGB as well as a relevant range of (initial) rotation velocities. 

Secondly, we note that the composite RCs are \emph{not} seen in eMSTO clusters
for which the ``pseudo-age'' distribution does not indicate any significant
number of stars with ages $\la 1.35$ Gyr, such as NGC 1651, NGC 1718, and NGC
2213. This suggests that the composite RCs are \emph{not} caused by interactive
binaries as proposed by \citet{li+12}, since the effect of interactive binaries
on the RC morphology is not expected to depend on age (in the age range 1\,--\,2
Gyr). 

\subsection{Comparison with Extended Star Formation Scenario}
\label{s:inter_sf}

In the context of the ``in situ star formation'' scenario, we first
note that Figure~\ref{f:Vescplot} suggests that the onset of the eMSTO
phenomenon occurs in the range $12 \la v_{\rm esc,\,7} \la 15$
\kms. This range of early escape velocities agrees well with observed
expansion velocities of the ejecta of the ``polluter'' stars  thought to
produce the Na-O anticorrelations among stars in globular clusters, as
detailed below.  

As to the case of IM-AGB stars, we turn our attention to OH/IR stars
featuring the superwind phase on the upper AGB, which is thought to
account for the bulk of mass loss of intermediate-mass stars
\citep[e.g.,][]{vaswoo93}.  
Radio observations of thermally pulsating OH/IR stars
in our Galaxy show expansion velocities $v_{\rm exp}$ in the range
14\,--\,21 \kms, peaking at $\simeq$\,17 \kms\
\citep[e.g.,][]{eder+88,tlh+91}. While expansion velocity
measurements of OH/IR stars in the Magellanic Clouds are still scarce,
4 LMC OH/IR stars have been found to exhibit $v_{\rm exp}$ values that
are $\sim$\,10\,--\,20\% lower than the Galactic ones in a given OH
luminosity class \citep{zijl+96}.  
Taking this ratio at face value, this would translate into $v_{\rm
  exp}$ values in the range 12\,--\,18 \kms\ for OH/IR stars in the LMC,
peaking at $\simeq$\,15 \kms. This is exactly the range of early
escape velocities within which we see the bifurcation between star
clusters with versus without eMSTOs.  

As to the case of massive stars, observations of nearby star forming
regions suggest that most such stars are in binary systems
\citep[e.g.,][]{sana+12}. Hence we focus on the case of massive binary
stars, for which imaging and spectroscopic observations have shown that the
enriched material is mainly ejected in a disc or ring geometry with expansion 
speeds in the range 15\,--\,50 \kms\
\citep[e.g.,][]{smit+02,smit+07}. Hence, the retention 
of mass-loss material from massive binary stars seems to require
somewhat higher cluster escape velocities than that from IM-AGB
stars. I.e., the rate of retention and accumulation of mass loss
material from massive binary stars may scale with the clusters'
(early) escape velocities, and this material may not be available in
significant quantities to eMSTO clusters with the lowest early escape
velocities. 

Next, we consider the hypothesis that the observed correlation between
MSTO width and $v_{\rm esc,\,7}$ reflects the (evolving) depth of the
potential well in the central regions of star clusters and its impact
on the ability of a star cluster to \emph{(i)} accrete an adequate amount
  of ``pristine'' gas from their surroundings, and/or \emph{(ii)} retain
chemically enriched material ejected by first-generation ``polluter'' stars, and
make the resulting material available for second-generation star formation. To
test this hypothesis, we use the cluster mass loss simulations described in 
\S\,\ref{s:dynevol} which provide cluster mass and $v_{\rm esc}$ as function of
time for the eMSTO clusters in our sample. 

\begin{figure*}[tbh]
%\centerline{\includegraphics[width=\textwidth]{Vesc_Time.ps}}
\centerline{\includegraphics[width=\textwidth]{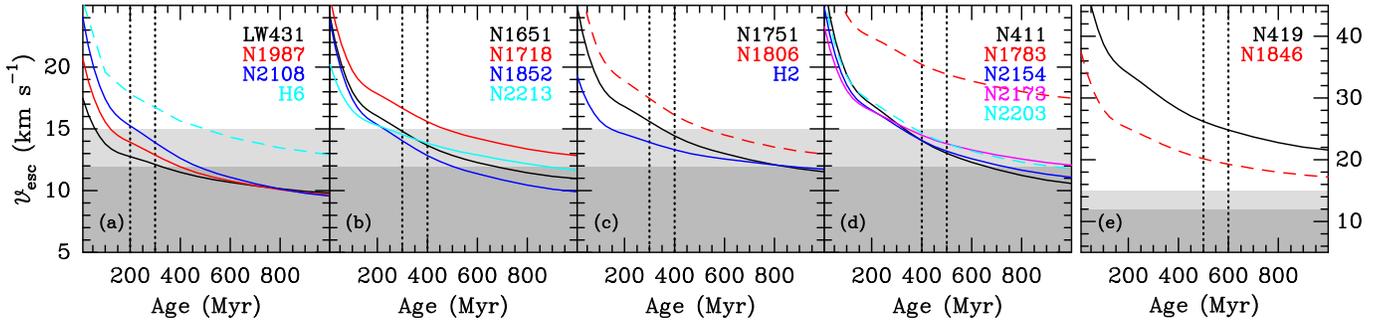}}
\caption{$v_{\rm esc}$ versus time since cluster birth. Each curve
  depicts dynamical evolution simulations for a given cluster,
  identified in the legend at the top right in each panel. Dashed
  lines identify clusters for which the assignment of a particular
  level of initial mass segregation is relatively uncertain (i.e., clusters
  shown with red circles in Figs.~\ref{f:Vescplot} and
  \ref{f:W20compare}). Light grey and dark grey areas indicate the 
  ranges $v_{\rm esc} \leq 15$ \kms\ and $v_{\rm esc} \leq 12$ \kms,
  respectively. Vertical dotted lines in each panel delineate the
  range in {\it FWHM}$_{\it MSTO}$ encompassed by the clusters drawn in that
  panel. Note that panel (e) has a different vertical scale than
  panels (a)\,--\,(d). See discussion in \S\,\ref{s:inter_sf}. 
}
\label{f:vesc_time}
\end{figure*}

As indicated by Table~\ref{t:dynamics}, the masses of the clusters in our
  sample at an age of 10 Myr ranged between roughly $1 \times 10^5\;M_{\odot}$
  and $1 \times 10^6$ $M_{\odot}$, for a reasonable range of levels of initial
  mass segregation\footnote{or between $\sim 6\times10^4\;M_{\odot}$ and
    $6\times10^5\;M_{\odot}$ for a Kroupa IMF.}. For such cluster
  masses, the calculations of \citet[][see their Figs.\ 2\,--\,3]{conspe11}
  predict that they were able to accrete $\approx$\,10\,--\,30\% of their mass
  in ``pristine'' gas unless the velocity of the cluster relative to that of the
  surrounding ISM was $\ga 600$ \kms. Such high relative velocities seem unlikely to
  occur in dwarf galaxies like the Magellanic Clouds (although it may well occur
  in the violent environment of merging massive galaxies). Even when taking into
  account the simplifying assumptions made in the study of \citet{conspe11}, it
  therefore seems plausible that the eMSTO clusters in our sample were able to
  accrete significant amounts of pristine gas from their surroundings. 
To test part \emph{(ii)} of the hypothesis mentioned above, we plot $v_{\rm
  esc}$ as function of time for the eMSTO clusters in our sample in  
Figures~\ref{f:vesc_time}a\,--\,\ref{f:vesc_time}e in which each panel shows
clusters with {\it FWHM}$_{\it MSTO}$ values in a given range. The values of
$v_{\rm esc}$ plotted in Figure~\ref{f:vesc_time} reflect the same levels  
of initial mass segregation as those used for the ``large'' symbols in 
Figure~\ref{f:Vescplot} (cf.\ \S\,\ref{sub:correl} above). 

Figures~\ref{f:vesc_time}a\,--\,\ref{f:vesc_time}d show that $v_{\rm
  esc}$ stays above 15 \kms\ for ages up to 100\,--\,150 Myr for \emph{all}
eMSTO clusters in our sample. This is equivalent to the lifetime of stars of
$\approx 4\;M_{\odot}$ \citep[e.g.,][]{mari+08}, and hence long enough for the
slow winds of massive binary stars and IM-AGB stars of the first generation to  
produce significant amounts of ``polluted'' material out of which
second-generation stars may be formed. 
This consistency between the escape velocity that seems to be required
for retention of enriched mass-loss material and the escape velocities
of eMSTO clusters at the time when the candidate polluter stars are present 
constitutes evidence in favor of the hypothesis stated above and hence
of the ``in situ star formation'' scenario, in which the {\it FWHM}$_{\it
  MSTO}$ values are a measure of the length of star formation activity. 

Among the eMSTO clusters in our sample with $200 \la {\it FWHM}_{\it
  MSTO}/{\rm Myr} \la 500$, $v_{\rm esc}$ crosses the range 12\,--\,15 \kms\
at ages similar to those indicated by their values of {\it FWHM}$_{\it
  MSTO}$. 
As shown by Figure~\ref{f:vesc_time}e, this behavior is not shared by
the clusters with the largest values of {\it FWHM}$_{\it MSTO}$ (i.e., in
the approximate range 500\,--\,550 Myr) in that their escape velocities stay
above 15 \kms\ for periods longer than that indicated by their values
of {\it FWHM}$_{\it MSTO}$. In the context of the ``in situ star formation''
scenario, this finding seems to indicate that the \emph{maximum}
length of the star formation era in the most massive star clusters is
\emph{not} set by the ability of a star cluster to retain chemically enriched
material from polluter stars and/or to accrete gas from the surrounding
  ISM. Instead, we speculate that the observational limit of {\it FWHM}$_{\it
  MSTO} \la 550$ Myr may reflect the typical time when the collective rate of
supernova (SN) events (i.e., ``prompt'' SN type Ia events by first-generation
stars plus any SN\,II events by second-generation stars) starts to be high
enough to sweep out the remaining gas in star clusters, thus ending the star
formation era \citep[see also][]{conspe11}\footnote{In this
  context, the observed values of {\it FWHM}$_{\it MSTO}$ in
  these clusters seem most consistent with the upper end of the published
  range of (a priori unknown) delay time scales $t_{\rm delay}$ for ``prompt''
  SN\,Ia explosions, which is $40 \la t_{\rm delay}/{\rm Myr} \la 400$
  \citep[e.g.,][]{mann+06,bran+10,maoz+10}.}. 

\begin{figure}[tbh]
%\centerline{\includegraphics[width=4.5cm]{n1856_n1866_Vesc.ps}}
\centerline{\includegraphics[width=4.5cm]{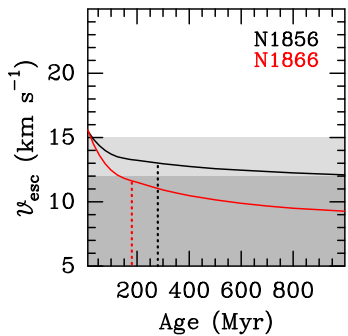}}
\caption{Similar to Figure~\ref{f:vesc_time}, but now for the young LMC clusters
  NGC 1856 and NGC 1866, using mass and radius data from 
  \citet{mclvdm05}. 
  Vertical dotted lines in this Figure indicate the current ages of the two
  clusters (280 Myr for NGC 1856 and 180 Myr for NGC 1866). See discussion in
  \S\,\ref{sub:comp_lit}.   
}
\label{f:n1856_1866}
\end{figure}

\subsection{Comparison with Recent Literature}
\label{sub:comp_lit}

Our results regarding the nature of eMSTO's in intermediate-age star clusters
generally favor the scenario in which eMSTO's reflect a range of stellar
  ages in the cluster rather than a range of stellar rotation velocities
among the MSTO stars. However, this conclusion seems to be at odds with a
number of recent studies that showed that younger star clusters do not exhibit
such age spreads. We discuss results of the latter studies in the context of
the ``in situ'' scenario below. Recapitulating the latter scenario, if massive
  binary stars and/or rapidly rotating massive stars are the main source of gas
  out of which a second generation is to be formed, one would expect that
  material to become available for star formation after 5\,--\,20 Myr of a
  cluster's life (if it is not swept out of the cluster by SN type II explosions
  of the massive stars). If instead AGB stars with $5 \la {\cal{M}}/M_{\odot} 
  \la 8$ are the main source of the enriched material, it would take
  $\sim$\,50\,--\,100 Myr to make it available for star
  formation. The actual era of second-generation star formation might not
  actually start until $\sim$\,100\,--\,150 Myr after the creation of the
  cluster, depending on the role of Lyman continuum photons from the massive
  stars of the first generation in prohibiting star formation through
  photodissociation of H$_2$ \citep{conspe11}. This era could last an
  additional few 10$^8$ yr depending on when the molecular gas reservoirs 
  run out or when the rate of ``prompt'' SN type Ia events by first-generation stars
  becomes significant.  

\begin{itemize}
\item
The recent review of \citet{port+10} presented known properties of young massive
clusters (YMCs) in our Galaxy, including a number whose escape velocity
is of order 12\,--\,15 \kms\ (e.g., the Arches cluster, NGC 3603, RSGC01,
RSGC03, and Westerlund 1). \citet{peri+09} also presented a study of van den
Bergh 0, a YMC with similar properties in M31. 
Such clusters would therefore be expected to be able to host extended
star formation if the ``in situ'' scenario is correct. However, the CMDs of these
YMCs do not show any signs of a significant spread in age. We suggest that these
observations can be reconciled with the AGB version of the ``in situ'' scenario
because the ages of all of these YMCs are $\la 25$ Myr. At that time, the slow
stellar winds from AGB stars have only just started, so that one would not yet
expect to see any significant sign of second-generation stars. 
\item
\citet{lars+11} presented CMDs of 6 massive YMCs ($10^5 - 10^6\;M_{\odot}$) in
galaxies nearby enough to resolve the outskirts of the YMCs with HST
photometry. The ages of these YMCs were found to be in the range 5\,--\,50
Myr. They find no evidence for significant age spreads, except for some of the
older YMCs which show tentative evidence of age spreads of up to 30 Myr. Given
\emph{(i)} the time it takes for a SSP to produce significant amounts of gas
from AGB mass loss and \emph{(ii)} the fact that the second-generation stars
are thought to form in the innermost regions of these clusters 
  (e.g., $\simeq$\,85\% of the second-generation stars in the
  \citet{derc+08} model are within 0.1 pc from the center, which region is
  overcrowded at the distances of these clusters), we believe these
  observations are not inconsistent with the ``in situ'' scenario. 
\item 
\citet{bassil13} studied two relatively massive
young LMC clusters (NGC~1856 and NGC~1866, with ages of 280 Myr and 
180 Myr, respectively) and found no evidence for age spreads larger than about
20\,--\,35 Myr, which they interpreted as a suggestion that the eMSTO feature in
intermediate-age clusters cannot be due to age spreads. To test this in the
context of the dynamical properties of these two clusters, we follow
\citet{bassil13} by adopting the King model fits of those two clusters from
the compilation of \citet[][see Table~1 in 
\citealt{bassil13}]{mclvdm05}, and multiply their masses by a factor 1.6
 since they were using the Chabrier IMF whereas we use the Salpeter IMF. After
running the dynamical evolution models described in \S\,\ref{s:dynevol} on
those two clusters, we plot their resulting $v_{\rm esc}$ as function of time
in Figure~\ref{f:n1856_1866}, whose setup is similar to that of
Figure~\ref{f:vesc_time}.  In choosing the plausible value for $v_{\rm
  esc,\,7}$ for these clusters, we recognize that the observed range
of core radii of Magellanic Cloud star clusters in the age range 200\,--\,300 
Myr is approximately 1.5\,--\,4.5 pc \citep{mack+08b} and hence we estimate  
$v_{\rm esc,\,7}^{p}$ as follows: 
\begin{equation}
v_{\rm esc,\,7}^{p} \equiv v_{\rm esc,\,7}^{\rm noseg} + (v_{\rm esc,\,7}^{\rm seg}
 - v_{\rm esc,\,7}^{\rm noseg})\; \times \left( \frac{\rc-1.5}{4.5-1.5} \right)\mbox{.}
\label{eq:plausible2}
\end{equation}
Comparing the solid lines in Figure~\ref{f:n1856_1866} with those in
\ref{f:vesc_time}, one sees that $v_{\rm esc}$ for NGC 1856 and NGC 1866 never
surpassed $\sim$\,15 \kms, whereas it did for all eMSTO clusters in our
sample. However, the difference between the $v_{\rm esc,\,7}^{p}$ of NGC 1856
and NGC 1866 and that of the lowest-mass eMSTO clusters in our sample is
relatively small (e.g., $v_{\rm esc,\,7}^{p} \simeq 17$ \kms\ for LW431),
indicating that the early escape velocity threshold that differentiates
clusters with eMSTOs from those without might occur close to 15 \kms\ when
assuming a Salpeter IMF \citep[see also][]{corr+14}. In that sense, the
apparent absence of eMSTOs in NGC 1856 and NGC 1866 is not necessarily 
inconsistent with the ``in situ star formation'' scenario, although the
margins seem to be small.  

In this context, we note that our King-model fits were done using
completeness-corrected surface number densities, whereas \citet{mclvdm05} used
surface brightness data to derive structural parameters for NGC 1856 and NGC
1866. The latter method is sensitive to the presence of mass segregation in
the sense that mass-segregated clusters will appear to have smaller radii (and
hence higher escape velocities) when using surface brightness data than when
using plain surface number densities. A study of the impact of this effect 
is currently underway using new \HST data of NGC 1856 (M.\ Correnti et al., in 
preparation). 
\item 
\citet{bast+13b} studied the presence of ongoing star formation in a large
sample of $\sim$\,130 YMCs, mainly using spectroscopy in the 4500\,--\,6000 \AA\ 
wavelength region, by means of H$\beta$ and {\sc [O\,iii]}$\lambda$5007
emission. Their sample of YMCs covered a significant range in best-fit ages
(10\,--\,1000 Myr) and masses ($10^4 - 10^8\;M_{\odot}$), both derived from
\emph{UBVRI} photometry or the spectra themselves. Concentrating on YMCs that
can significantly constrain the AGB version of the ``in situ'' scenario, we
select YMCs from their sample that have: \emph{(i)} ${\cal{M}} \ga
10^5\;M_{\odot}$ (to create a high probability that $v_{\rm esc} \ga 15$
\kms), \emph{(ii)} ages in the range 100\,--\,300 Myr, and \emph{(iii)}
spectra that are shown in the literature (either in \citet{bast+13b} itself or
in the references therein). This selection results in a sample of 21
YMCs. Inspection of their spectra reveals 4 clusters that seem to show 
  hints of {\sc [O\,iii]}$\lambda$5007 in emission and/or H$\beta$ emission
filling in the deep absorption line (clusters M82-43.2, M82-98, NGC3921-S2,
and NGC2997-376), i.e., $\la$\,20\% of the sample.  

This apparent lack of emission in a significant fraction of this subsample of
YMCs studied by \citet{bast+13b} provides an important constraint to the
``in situ'' scenario, and therefore merits some discussion. One relevant
consideration may be that the second-generation star formation is thought to
occur in the innermost regions of the clusters
(likely in a flattened structure), and extinction by the ISM in
those regions may impact the detection of line emission. This possibility can
be tested in the future by performing spectral observations at longer
wavelengths and with emission lines that are intrinsically stronger than
H$\beta$ and {\sc [O\,iii]}$\lambda$5007 in {\sc H\,ii} regions (e.g.,
H$\alpha$ and Br$\gamma$ from the ground, Pa$\alpha$ and Br$\alpha$ from
space). Another consideration is that the duration of the line emission era in
star-forming regions is only $\simeq$\,7 Myr, which is a very small fraction of
the age spread indicated by the width of eMSTOs (or the time interval in which
second-generation star formation is predicted to occur in the ``in situ''
scenario). It is not known whether this star formation would be occurring in a
continuous fashion or perhaps in recurrent episodes. While the pseudo-age
distributions of the clusters in our sample
(i.e,. Figs.~\ref{f:cmdplot1}\,--\,\ref{f:cmdplot3}) do typically appear quite 
smooth, suggesting continuous star formation activity, we remind the
reader that our time resolution element is similar to a gaussian with
FWHM~$\simeq$~150\,--\,200 Myr, depending on the cluster age. We therefore
cannot detect variations in the age distribution on time scales of several tens
of Myr, leaving open the possibility of recurrent star formation activity. 
Finally, significant line emission in star-forming regions only occurs when
there are enough O and B stars to ionize the gas, implying a dependence on the
(a priori unknown) IMF of the second generation. The apparent lack of line
emission in several YMCs at ages of $\sim$\,100\,--\,300 Myr might therefore
still be consistent with the ``in situ'' scenario if second-generation stars
are formed in regions where the IMF is such that O and B stars are relatively
unlikely to form, and/or where the SFR is small relative to that of
first-generation star-forming regions. We suggest that the likelihood of these
possibilities be tested in the near future using new simulations as well as
(IR) observations. 
\item
\citet{cabr+14} studied the SFH of the very massive young Cluster \#1 in the
merger remnant galaxy NGC 34, using a spectrum obtained earlier by
\citet{schsei07}. Using stellar population synthesis fitting, they find that
the SFH of this cluster is consistent with a SSP of age 100 $\pm$ 30 Myr and
rule out the presence of a second population that is younger than 70 Myr at a
second-to-first-generation mass ratio of $\geq 0.1$. While these results
provide important constraints on the presence of a second stellar generation 
in this cluster, it is not clear yet how much second-generation star formation
one might expect at a cluster age of 100 Myr. According to \citet{conspe11},
the density of Lyman continuum photons from massive stars of the first
generation would likely still be high enough at this age to prohibit new star
formation. Similar work for massive clusters at ages of 200\,--\,500 Myr should 
therefore yield more relevant constraints to the ``in situ'' scenario. 
\end{itemize}

\subsection{Comparison with Other Scenarios Involving a Range of Stellar Ages}
\label{sub:other}

The presence of a range of stellar ages within a star cluster does not
necessarily imply that all cluster stars were formed ``in situ'' within
the clusters. In this context, we briefly comment on the feasibility of two
scenarios that involve an \emph{external} origin of part of the stars in
clusters while preserving the observed homogeneity in [Fe/H]: (1) the merger
of two or more star clusters formed in a given giant molecular cloud
(hereafter GMC), and (2) the merger of a (young) star cluster with a GMC.    

As to possibility (1) above, a range of ages in clusters could be the result
of merging of smaller clusters that were all formed by the collapse of a given 
GMC (in which multiple clusters were formed at different times). 
However, as explained by \citet{goud+09}, the observed age ranges of
200\,--\,500 Myr in eMSTO clusters are much larger than the observed age 
differences between binary or multiple clusters in the LMC
\citep[e.g.,][]{dieb+02}. Hence, it seems hard to form the eMSTO clusters by
star formation within a given GMC in general, especially since the eMSTO
phenomenon is very common among intermediate-age clusters in the Magellanic
Clouds.  

As to possibility (2), the simulations by \citet{bekmac09} suggest that new
episodes of star formation can be triggered by an interaction of a star
cluster with a GMC, as long as the space velocity of the star cluster 
relative to that of the GMC is smaller than $\sim$\,2 times the internal
velocity dispersion of the cluster. 
Several aspects of this scenario seem to be generally consistent with the
observational evidence: \\ [-2.5ex]
\begin{itemize}
\item 
As argued by \citet{bekmac09}, the typical time scale
of a cluster-GMC merger can be of the same order as that indicated by the
MSTO widths of eMSTO clusters if the (average) surface number density of GMCs
in the LMC was a few times higher than it is now.  
This does not seem implausible: 
  According to the SFH of the LMC published by \citet{weis+13}, 
  the LMC formed $\simeq$\,25\% of its current stars over the last 2
  Gyr. 
  Using the current stellar and gas masses of the LMC given by
  \citet{vdm+02}, this implies that the gas supply of the LMC decreased by 
  $0.25 \; {\cal{M}}_{\ast,\, {\rm LMC}} = 7.6 \times 10^8\; M_{\odot}$
  over the last 2 Gyr. This is about 1.5 times its current
  gas mass, so that the LMC gas supply was a factor $\approx$\,2.5 larger
  when the clusters in our sample were created. 
  Furthermore, the era of 1\,--\,2 Gyr ago in the Magellanic
  Clouds is thought to feature strong tidal interactions between the LMC and
  SMC, causing strong star (and star cluster) formation in the bar and NW arm
  of the LMC \citep[e.g.,][]{bekk+04,diabek11,besl+12,rube+12,piat14}, where
  many of the clusters in our sample are located. It thus seems reasonable to
  postulate that relatively high number densities of high-mass GMCs were
  relatively common at the time, allowing the formation of several massive
  star clusters and possibly creating a situation where the typical time scale
  of cluster-GMC mergers was similar to the age ranges
  indicated by {\it FWHM}$_{\it MSTO}$ values of eMSTO clusters. 
  Alternatively, the time scale of $\sim$\,100\,--\,500 Myr may reflect the
  typical life time of strong density waves during the tidal interactions
  between the LMC and SMC at the time, allowing strong cluster formation and
  cluster-GMC interactions to occur during that time period.  \\ [-3.3ex]
\item 
The correlations between {\it FWHM}$_{\it MSTO}$ and cluster escape velocity
and mass shown in \S\,\ref{sub:correl} above also seem to be 
consistent with this scenario in that more massive pre-existing (``seed'')
clusters should be capable to merge with more (and more massive) GMCs relative
to less massive seed clusters, which would allow the sampling of a wider 
range of stellar ages. 
\end{itemize}
Note that the cluster-GMC merger scenario would in principle also predict the 
existence of eMSTO clusters with {\it FWHM}$_{\it MSTO} \ga 1$ Gyr, namely if
local number densities of massive GMCs are similar to the average current
value \citep{bekmac09}. However, such eMSTO clusters are not observed. 
Reconciling this in the context of this scenario would require the time scale 
of gas consumption by star formation to be short enough to render the number
density and size of GMCs to be too small to produce significant cluster-GMC 
mergers after $\sim$\,1 Gyr. This is consistent with the dip seen in the
average SFHs of the LMC and SMC between lookback times of $\sim$\,0.5 and 1
Gyr, after a period of strong star formation between 1 and 2.5\,--\,3 Gyr ago
\citep{weis+13}. 

We conclude that the cluster-GMC merger scenario of \citet{bekmac09} can in
principle explain many observed properties of eMSTO clusters and provides a 
relevant alternative to the ``in situ star formation'' scenario.

\subsection{Comparison with Light-Element Abundance Variations in LMC
  Clusters}
\label{sub:NaFe}

If eMSTO clusters and ancient Galactic GCs share a formation process that
involves star formation over a time span of a few 10$^8$ yr, an important
prediction would be that eMSTO clusters ought to show some level of
star-to-star variations in light-element abundances in a way similar to the
Na-O anti-correlations seen among Galactic GCs. Conversely, if the latter are
mainly due to enrichment by winds of massive (binary) stars, which generally
feature higher wind speeds than do AGB stars, the amplitude of light-element
abundance variations in eMSTO clusters would be expected to be lower or even
negligible, since it is likely that the Galactic GCs that show the Na-O
anti-correlation were significantly more massive at birth than the eMSTO
clusters in the Magellanic Clouds.  

In this section, we attempt to estimate the expected amplitude of
light-element abundance variations in the eMSTO clusters in our sample and
compare our estimate with the available data. 

As shown by several studies, the Na-O anti-correlation in Galactic GCs can be 
  reproduced with a simple ``in situ'' model in which second-generation stars
  are formed from pristine and processed material mixed in varying amounts
  \citep{pran+07,vendan08,derc+10,conr12}. In this context, ``processed''
  material has enhanced [Na/Fe] and depleted [O/Fe] relative to
  ``pristine'' material. 
One important feature of the Na-O anti-correlation among stars in individual
GCs is that its extent in a [Na/Fe] versus [O/Fe] diagram correlates 
with cluster mass \citep{carr+07,carr+10}. As shown by
\citet{conr12}, this trend is consistent with a simple dilution scenario such
as that mentioned above if the Galactic GCs lost of order 90\% of their
initial mass during their life time.   

In the context of the current exercise, we adopt the values of $f_{\rm p}$,
the fraction of the GC mass made from pure processed material, in Galactic GCs
from the study of \citet{conr12}. To create predictions for $f_{\rm p}$ in the 
clusters in our sample (which have ages of 1\,--\,2 Gyr), we estimate the
masses that the Galactic GCs in the sample of \citet{conr12} would have had
at an age of 2 Gyr. In doing so, we make the assumptions that \emph{(i)} Galactic
GCs have a current age of 13 Gyr and \emph{(ii)} the mass loss rate of
Galactic GCs between the ages of 2 and 13 Gyr was dominated by long-term
disruption mechanisms such as two-body relaxation. Using effective radius data
from the 2010 update of the catalog of \citet{harr96}, we then apply the
mass-density-dependent mass loss rates of \citet[][their eq.\ 5]{mclfal08} to
yield estimates for ${\cal{M}}_{\rm GC,\,2}$, the masses of the Galactic GCs
at an age of 2 Gyr. The relevant properties of these Galactic GCs are listed
in Table~\ref{t:GGCs}.

% Place Table 5 here
\begin{table}[tbh]
\begin{center}
\footnotesize
%\scriptsize
\caption{Properties of Galactic GCs.}
 \label{t:GGCs}
%\bf
\begin{tabular}{@{}lcrcccc@{}}
\multicolumn{3}{c}{~} \\ [-2.5ex]   
 \tableline \tableline
\multicolumn{3}{c}{~} \\ [-1.8ex]                                                
\multicolumn{1}{c}{Cluster} & \multicolumn{1}{c}{[Fe/H]} & N$_{\ast}$ &
 log\,${\cal{M}}_{\rm GC}$ & $f_{\rm p}$ & $r_{\rm e}$ & log\,${\cal{M}}_{\rm
   GC,\,2}$   \\
\multicolumn{1}{c}{(1)}     & \multicolumn{1}{c}{(2)}     & (3) & (4) & (5) &
 (6) & (7) \\ [0.5ex] \tableline  
\multicolumn{3}{c}{~} \\ [-1.5ex]              
%   NGC       [Fe/H]    N     logM    fp      r_e    logM_2
  NGC 7099  & $-$2.33 &  19  & 5.19 &  0.32 & 1.26 &  5.96 \\
  NGC 7078  & $-$2.33 &  20  & 5.89 &  0.36 & 1.88 &  6.23 \\
  NGC 4590  & $-$2.23 &  36  & 5.16 &  0.28 & 7.03 &  5.30 \\
  NGC 6397  & $-$1.98 &  13  & 4.87 &  0.20 & 1.56 &  5.66 \\
  NGC 6809  & $-$1.98 &  75  & 5.24 &  0.33 & 5.70 &  5.41 \\
  NGC 6715  & $-$1.57 &  76  & 6.23 &  0.42 & 6.40 &  6.28 \\
  NGC 1904  & $-$1.55 &  39  & 5.37 &  0.31 & 2.45 &  5.76 \\
  NGC 6752  & $-$1.56 &  88  & 5.31 &  0.36 & 1.78 &  5.86 \\
  NGC 6254  & $-$1.56 &  77  & 5.21 &  0.27 & 2.88 &  5.59 \\
  NGC 3201  & $-$1.50 &  94  & 5.21 &  0.34 & 5.06 &  5.41 \\
  NGC 5904  & $-$1.34 & 106  & 5.75 &  0.38 & 3.85 &  5.92 \\
  NGC 6218  & $-$1.31 &  66  & 5.15 &  0.34 & 2.67 &  5.57 \\
  NGC 288   & $-$1.23 &  64  & 4.92 &  0.29 & 5.49 &  5.16 \\
  NGC 6121  & $-$1.20 &  80  & 5.10 &  0.33 & 2.77 &  5.52 \\
  NGC 6171  & $-$1.06 &  27  & 5.07 &  0.31 & 2.99 &  5.47 \\
  NGC 2808  & $-$1.10 &  90  & 5.98 &  0.42 & 2.79 &  6.18 \\
  NGC 6838  & $-$0.80 &  31  & 4.46 &  0.25 & 1.53 &  5.28 \\
  NGC 104   & $-$0.74 & 109  & 5.99 &  0.40 & 4.15 &  6.11 \\
  NGC 6388  & $-$0.40 &  29  & 5.99 &  0.39 & 1.52 &  6.38 \\
  NGC 6441  & $-$0.34 &  24  & 6.08 &  0.36 & 1.95 &  6.36 \\ [0.5ex] \tableline
\multicolumn{3}{c}{~} \\ [-2.5ex]              
\end{tabular}
\tablecomments{Columns: (1): GC ID; (2): [Fe/H] in dex; (3): Number of
  stars used in the abundance analysis (see \citealt{conr12}); (4): Log of
  current GC mass in $M_{\odot}$. (5): mass fraction of processed material ;
  (6): Effective radius from \citet{harr96} in pc; (7): log of GC mass
  at age of 2 Gyr in $M_{\odot}$. }
\end{center}
\end{table}

\begin{figure}[tbh]
%\centerline{\includegraphics[width=8cm]{f11.ps}}
\centerline{\includegraphics[width=7.5cm]{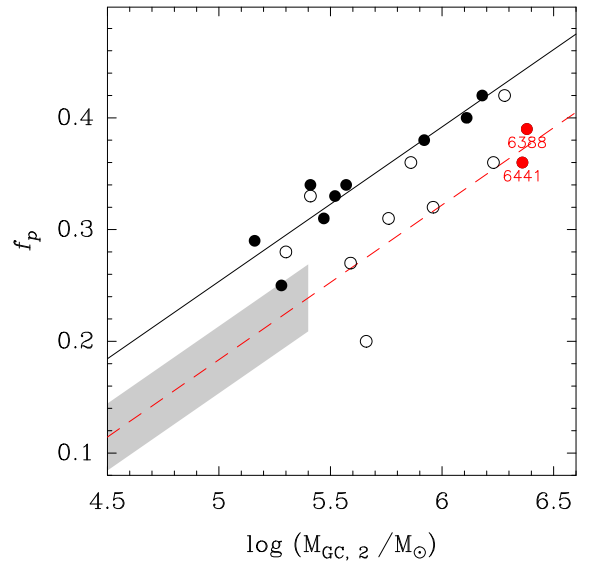}}
\caption{The fraction of GC mass comprised of pure processed material,
  $f_{\rm p}$, versus GC mass at an age of 2 Gyr for the Galactic GCs in the
  sample of \citet{conr12}. Filled circles represent GCs with $\mbox{[Fe/H]} >
  -1.5$, whereas open circles represent more metal-poor GCs. The two most
  metal-rich GCs in this sample (NGC 6388 and NGC 6441) are shown in red and
  labeled. The solid line represents a best-fit linear relation for the GCs
  with $\mbox{[Fe/H]} > -1.5$ \emph{except} NGC 6388 and NGC 6441. The dashed
  line is the same as the solid line after shifting it down by 0.07 dex. An
  estimate of the area expected to be populated by the Magellanic Cloud
  clusters in our sample is shown in grey. See discussion in \S\,\ref{sub:NaFe}. 
}
\label{f:f_p_plot}
\end{figure}

Figure~\ref{f:f_p_plot} shows $f_{\rm p}$ as a function of
${\cal{M}}_{\rm GC,\,2}$. Different symbols represent GCs with different
[Fe/H]. As reported by \citet{conr12} for \emph{current} GC masses, there is a strong
linear correlation between $\log\,({\cal{M}}_{\rm GC,\,2})$ and $f_{\rm p}$, which is
generally stronger for the ``metal-rich'' Galactic GCs (those with [Fe/H] $>
-1.5$) than for the metal-poor ones. However, the two most metal-rich Galactic GCs
in the compilation of \citet{conr12} (i.e., NGC 6388 ([Fe/H] = $-$0.40) and NGC
6441 ([Fe/H = $-$0.34)) turn out to feature $f_{\rm p}$ values that are
systematically below the relation defined by the other ``metal-rich'' GCs (by
$\simeq 0.07 \; \mbox{dex} \simeq 5\sigma$). We suggest that this is a
manifestation of the metallicity dependence of Oxygen yields in AGB
models. The recent models of \citet{vent+13} show this quite clearly. In their
$Z = 0.008$ models, the mean [O/Fe] yield for AGB stars with masses in the
range 4\,--\,8 $M_{\odot}$ is  0.00 $\pm$ 0.02. At metallicities such as
those of NGC 6388 and NGC 6441, the full range of [O/Fe] is therefore expected to 
be near zero, while this range commonly exceeds 1 dex for high-mass
low-metallicity GCs \citep[see, e.g.,][]{carr+10,conr12}. Note however that in
contrast with [O/Fe], the predicted range of [Na/Fe] in $Z = 0.008$ models is
similar to that of lower-metallicity models \citep{vent+13}. Hence, it seems
fair to postulate that the relatively low values of $f_{\rm p}$ for NGC 6388
and NGC 6441 are due to a relative lack of variation in [O/Fe]. Since $Z =
0.008$ is also the metallicity of the LMC clusters, one might expect their
$f_{\rm p}$ to be similarly low relative to the trend with mass defined by the
``metal-rich'' Galactic GCs (after removing NGC 6388 and NGC 6441). 

The grey area shown in Figure~\ref{f:f_p_plot} depicts the expected $f_{\rm
  p}$ values for the clusters in our sample (with $4.5 \la \log\,({\cal{M}}_{\rm
  cl}/M_{\odot}) \la 5.4$), under the assumptions mentioned above and allowing
for measurement uncertainties similar to those of the Galactic GCs. For a
relatively massive cluster in our sample with $5.0 \la \log \,({\cal{M}}_{\rm
  cl}/M_{\odot}) \la 5.4$, one would then expect $f_{\rm p}$ to be in the
approximate range 0.18\,--\,0.24 \emph{in case star formation occurred in situ
  in these clusters}. (Note that if cluster-GMC merging occurred in the early
life of these clusters, the seed clusters would have had a lower mass than in
the ``in situ'' case, so that the expected $f_{\rm p}$ values would be
smaller.) The two Galactic GCs in the sample of \citet{conr12} with $f_{\rm
  p}$ values in this approximate range are NGC 6397 ($f_{\rm p} = 0.20$) and
NGC 6838 ($f_{\rm p} = 0.25$). Defining $\Delta$\,[Na/Fe] as the FWHM of a
Gaussian fit to the distribution of [Na/Fe] of RGB stars in a given cluster,
the data in Table~9 of \citet{carr+09} yield $\Delta$\,[Na/Fe] = 0.38 dex for
both NGC 6397 and NGC 6838. We suggest that this is a suitable estimate for an
upper limit of $\Delta$\,[Na/Fe] in the clusters in our sample.  

\begin{figure}[tb]
%\centerline{\includegraphics[width=8cm]{f12.ps}}
\centerline{\includegraphics[width=7.5cm]{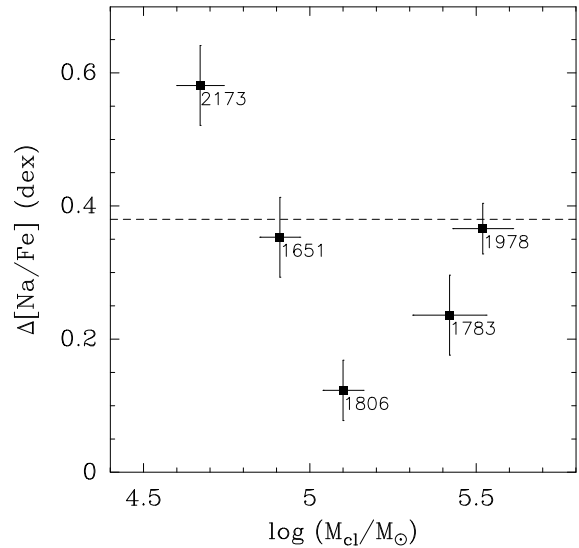}}
\caption{The variation in [Na/Fe] versus cluster mass for intermediate-age
  LMC clusters. [Na/Fe] data are from \citet{mucc+08,mucc+14}. NGC numbers of
  the clusters are labeled next to their data values. The dashed line
  illustrates $\Delta$\,[Na/Fe] = 0.38, the estimated upper limit of
  $\Delta$\,[Na/Fe] for such clusters. See discussion in \S\,\ref{sub:NaFe}. 
}
\label{f:NaFeplot}
\end{figure}

The currently available data on $\Delta$\,[Na/Fe] for intermediate-age clusters
in the Magellanic Clouds consist of elemental abundance measurements of 35
RGB stars in 5 LMC clusters, covering 5\,--\,11 stars per cluster
\citep{mucc+08,mucc+14}. Four of their five clusters are members of our
sample. We approximate $\Delta$\,[Na/Fe] for these clusters by means of the
FWHM of Gaussian fits to the distributions of [Na/Fe]. The measurement
uncertainty of $\Delta$\,[Na/Fe] is approximated by $\sigma_{\rm
  meas}/\sqrt(N_{\ast}-1)$ where $\sigma_{\rm meas}$ is the typical
measurement uncertainty of [Na/Fe] of single stars and $N_{\ast}$ is the
number of stars measured in a given cluster. Figure~\ref{f:NaFeplot} shows
$\Delta$\,[Na/Fe] as a function of the cluster mass. For NGC 1978, we estimate
its mass from the compilation of integrated-light 2MASS photometry by
\citet{pess+06}, in conjunction with the age and foreground reddening reported
by \citet{mucc+07} and ${\cal{M}}/L$ data from \citet{bc03} for a Salpeter IMF. 

Figure~\ref{f:NaFeplot} shows that the currently available $\Delta$\,[Na/Fe]
values of the intermediate-age clusters in the Magellanic Clouds range between
about 0.1 and 0.6 dex. While these values do not generally seem inconsistent
with the expectation based on the cluster mass dependence of $f_{\rm p}$ among
Galactic GCs described above, the situation is not yet clear given the 
significant scatter of $\Delta$\,[Na/Fe] among the clusters and the absence of
an obvious dependence on cluster mass. It is not clear to what extent this
scatter is caused by the small numbers of stars with spectroscopic abundance 
measurements in these clusters.

Given the importance of a statistically significant inventory of
light-element abundance variations in eMSTO clusters in terms of its relevance
in the context of formation scenarios of star clusters, it is important to
expand the effort of obtaining high-quality spectroscopic measurements of a
statistically adequate number of RGB stars in several eMSTO clusters,
with the target stars covering a suitable range of distance from the cluster
centers. The latter aspect is important since the age of eMSTO clusters is
often similar to their half-mass relaxation time, so that one would expect to
see differences in the radial distributions of stars of different age
\citep[see][]{goud+11b}.

\section{Concluding Remarks}  
\label{s:conc}

In an effort to further our understanding of the nature and demography of the
eMSTO phenomenon in intermediate-age star clusters in the Magellanic Clouds, we
have obtained new deep two-color imaging for 8 such clusters, using the WFC3
instrument aboard \emph{HST}. We combined the new data with data already
available in the \HST archive to establish high-quality photometry with the
ePSF fitting technique for a complete sample of 18 Magellanic Cloud star
clusters with integrated magnitude $V_{\rm tot}^0 < 12.5$ and integrated-light
colors that indicate ages between 1 and 2 Gyr.
The star clusters in our sample cover a range in present-day mass from about
$3\times 10^4$ \Msun\ to $4 \times 10^5$ \Msun.  
We used isochrones from the Padova family to determine best-fit population
parameters for all clusters in our sample, and we evaluated masses and
structural parameters for the clusters using \citet{king62} model fits to the
radial distribution of surface number densities. 
Using Monte-Carlo simulations, we created artificial CMDs for each cluster,
showing its morphology if it were a pure SSP (including unresolved binary
stars) for comparison with the observed CMDs. Finally, we evaluated central 
escape velocities ($v_{\rm esc}$) of the clusters as a function of time using
dynamical evolution calculations with and without initial mass segregation.  
Our main conclusions are the following. \\ [-2.5ex]
\begin{enumerate}
\item \emph{All} star clusters in our sample with ages in the range
  1\,--\,2~Gyr feature eMSTOs, i.e., MSTO regions that are wider than can be
  accounted for by a SSP. 
  FWHM widths of pseudo-age distributions derived from the eMSTO morphology are
  found to be equivalent to age spreads of 200\,--\,550 Myr. In contrast,
  similar data of two lower-mass star clusters in the same age range reveals
  significantly narrower MSTOs whose widths are consistent with those of their
  respective SSP simulations. \\ [-3.3ex] 
\item Star clusters featuring eMSTOs and whose pseudo-age distributions
  indicate the presence of significant numbers of stars in the age range
  1.0\,--\,1.3 Gyr also feature composite red clumps in their CMDs, even though
  their formal best-fit age is almost always older than 1.3 Gyr. Conversely,
  star clusters with eMSTOs but without significant numbers of stars with ages
  $\leq$\,1.3 Gyr do \emph{not} show composite red clumps. This constitutes
  evidence that eMSTOs are caused by a range of ages rather than a range of
  stellar rotation velocities or the presence of interacting binaries.  \\ [-3.3ex]
\item We find that $v_{\rm esc} \ga 15$ \kms\ out to ages of at least 100 Myr
  for \emph{all} clusters that feature eMSTOs, while $v_{\rm esc} \la 15$
  \kms\ at all ages for the two lower-mass clusters that do \emph{not} show
  eMSTOs. 
   In the context of the ``in situ star formation'' scenario, the eMSTO
    phenomenon would only occur in  
  clusters that feature early $v_{\rm esc}$ values that are higher than the
  wind velocities of the types of stars that have been proposed to provide the
  material from which second-generation stars can form. 
  Our result would then suggest 
  that the lower limit to such wind velocities is of order 15 \kms. This
  hypothesis is found to be consistent with observed wind velocities of
  intermediate-mass AGB stars and massive binary stars in the literature. It
  is also found to be consistent (albeit possibly only marginally) with
  the absence of eMSTOs in two young star clusters (with ages $\la 300$ Myr)
  that was recently reported by \citet{bassil13}. \\ [-3.3ex]  
\item We find a significant correlation between the {\it FWHM}$_{\it MSTO}$,
  the width of the pseudo-age distributions of eMSTO clusters, and their
  central escape velocity at an age of 10 Myr, $v_{\rm esc,\,7}$. This
    correlation persists when plotting {\it FWHM}$_{\it MSTO}$ versus
    \emph{current} central escape velocity, albeit at lower significance.  
    Similar correlations are found between {\it FWHM}$_{\it MSTO}$ and
    cluster mass as well.  
  We find that these correlations cannot be reproduced by the effects of a range
  of stellar rotation velocities within star clusters according to recent
  models. In particular, the observed MSTO widths of eMSTO clusters are larger
  than those predicted by the stellar rotation models, especially for the
  clusters with the larger values of $v_{\rm esc,\,7}$. Furthermore, it is not
  clear how to explain the absence of eMSTOs in the two lower-mass clusters 
  in the stellar rotation scenario. 
  We therefore argue that the eMSTO phenomenon among intermediate-age star
  clusters is mainly caused by extended star formation within the cluster, 
  likely from material shed by first-generation stars featuring slow
  stellar winds and/or chemically pristine material accreted from the
  ambient ISM at early times.  
\end{enumerate}

The overall general picture on the formation process of intermediate-age star
clusters featuring eMSTOs that seems to be most consistent with our results 
is as follows. The masses and central escape velocities of eMSTO clusters
in the first few 10$^8$ yr seem to have been high enough to accrete a
significant amount of ``pristine'' gas from the surroundings (by slow
 accretion and/or by merging with GMCs \`a la \citealt{bekmac09}) and/or 
retain a significant fraction of the gas supplied by slow winds of
``polluters'' (IM-AGB stars and massive binary stars) of the first generation,
and accumulate this material at the bottom of the clusters' potential wells,
making it available for secondary star formation. (This is not the case for
clusters whose central escape velocities never exceeded about 12 \kms.) During
the first few hundreds of Myr, clusters with higher escape velocities
generally seem to have been able to extend the star formation process for a
longer time than clusters with lower escape velocities, possibly by means of
ongoing accretion of pristine gas from the ambient ISM, merging with
  GMCs, and/or retention of enriched wind material
from newly formed polluter stars. The star formation era terminated when at
least one of two things occured: 
(a) the gas swept up and/or accumulated by the cluster is exhausted by
  star formation, or (b) the collective rate of SN events (i.e., ``prompt''
SN Ia events by first-generation stars and SN II events by second-generation
stars) started to be high enough to sweep out the remaining gas in star
clusters.  

If eMSTO clusters and ancient Galactic GCs share a formation process that
involves star formation over a time span of a few 10$^8$ yr, a key prediction
would be that eMSTO clusters should show some level of light-element abundance
variations in a way similar to the Na-O anti-correlations seen among Galactic
GCs.  
We estimated the expected level of
[Na/Fe] variations ($\Delta$\,[Na/Fe]) in eMSTO clusters by evaluating the
masses of Galactic GCs at an age of 2 Gyr by inversely applying their modeled
mass-loss rates during the last $\simeq$\,11 Gyr, followed by an extrapolation
of the correlation of the observed extents of the Na-O anti-correlations
within the GCs with their masses at an age of 2 Gyr. The estimated levels of
$\Delta$\,[Na/Fe] in the eMSTO clusters are found to be broadly consistent
with the currently available spectroscopic data, although there is a significant
scatter of $\Delta$\,[Na/Fe] among the clusters. It is not clear to what
extent this scatter is caused by the small numbers of stars with abundance
measurements in these clusters, and we urge the community to expand the effort
of obtaining high-quality spectroscopic measurements of a statistically
adequate number of RGB stars in several eMSTO clusters, with the target stars
covering a suitable range of distance from the cluster centers, thus making
sure that stars of different generations (if present) are sampled adequately. 
Several teams are presently pursuing such spectroscopic studies, whose
results are eagerly awaited.

\acknowledgments
We acknowledge useful discussions with Nate Bastian, Enrico Vesperini, and
Wuming Yang.    
We appreciated the efforts and comments of the anonymous referee, 
 which resulted in an improved paper. 
We made significant use of the SAO/NASA Astrophysics Data System during this
project. 
We thank Jay Anderson for his help in various aspects regarding the use of his
ePSF package.
Support for this project was provided by NASA through grant HST-GO-12257 from
the Space Telescope Science Institute, which is operated by the Association of 
Universities for Research in Astronomy, Inc., under NASA contract NAS5--26555.  
We acknowledge the use of the R Language for Statistical Computing, see
http://www.R-project.org.

Facilities: \facility{HST (ACS)}, \facility{HST (WFC3)}

%% If you are not including electonic art with your submission, you may
%% mark up your captions using the \figcaption command. See the 
%% No more than seven \figcaption commands are allowed per page, 
%% so if you have more than seven captions, insert a \clearpage 

%% Tables should be submitted one per page, so put a \clearpage before

%% Two options are available to the author for producing tables:  the
%% deluxetable environment provided by the AASTeX package or the LaTeX
%% table environment.  Use of deluxetable is preferred.
%%

%% Three table samples follow, two marked up in the deluxetable environment,
%% one marked up as a LaTeX table.

%% In this first example, note that the \tabletypesize{}
%% command has been used to reduce the font size of the table.
%% Note also that the \label command needs to be placed 

%% Tables may also be prepared as separate files. See the accompanying
%% sample file table.tex for an example of an external table file.
%% To include an external file in your main document, use the \input
%% command. Uncomment the line below to include table.tex in this
%% sample file. (Note that you will need to comment out the \documentclass,
%% \begin{document}, and \end{document} commands from table.tex if you want
%% to include it in this document.)

%\input{tab1.tex}

%\clearpage

%\input{tab2.tex}

%% The following command ends your manuscript. LaTeX will ignore any text

\clearpage
\begin{turnpage}
\setcounter{table}{2}
\begin{deluxetable}{lccc@{\hskip 3pt}rrc@{\hskip 3pt}rccccc@{}}
%\rotate
\tablewidth{0pt}
\tabletypesize{\scriptsize}
\tablecolumns{13}
\tablecaption{Derived dynamical parameters of star clusters in our full sample.  
\label{t:dynamics}}
\tablehead{
 \colhead{} & \multicolumn{3}{c}{log (${\cal{M}}_{\rm cl}/M_{\odot}$)} &
 \multicolumn{3}{|c}{$r_{\rm eff}$} & \multicolumn{4}{c}{\ } & \multicolumn{2}{c}{MSTO widths} \\
\cline{2-4}  \cline{5-7} \cline{12-13}
 \colhead{Cluster} & \colhead{Current} & \colhead{10 Myr} & \colhead{10 Myr, seg.} &
 \multicolumn{1}{|c}{Current} & \multicolumn{1}{c}{10 Myr} & \colhead{10 Myr, seg.} & 
 \multicolumn{1}{c}{$v_{\rm esc}$} & \colhead{$v_{\rm esc,\,7}^{\rm noseg}$} & 
 \colhead{$v_{\rm esc,\,7}^{\rm seg}$} & 
 \colhead{$v_{\rm esc,\,7}^{p}$} & \colhead{{\it FWHM}} & \colhead{{\it W20}} \\ 		
 \colhead{(1)} & \colhead{(2)} & \colhead{(3)} & \colhead{(4)} &
 \multicolumn{1}{c}{(5)} &  \multicolumn{1}{c}{(6)} & \colhead{(7)} & \multicolumn{1}{c}{(8)} &
 \colhead{(9)} & \colhead{(10)} & \colhead{(11)} & \colhead{(12)} & \colhead{(13)} 
}
%\multicolumn{1}{c}{(1)}     & \multicolumn{1}{c}{(2)}        & (3) & (4) & 
% (5) \\ [0.5ex] \tableline
\startdata                                  
%         
% NGC           LogM   err     logM7    err       logM7s   err        Re   err          Re7   err         Re7s   err       Vesc   err      Vesc7   err       Vesc7s  err          theVesc7  err           FWHM  W20
 NGC 411 &  4.67 $\pm$ 0.03 &  4.82 $\pm$ 0.03 &  5.24 $\pm$ 0.03 &  6.1 $\pm$ 0.8 &  4.8 $\pm$ 0.6 &  3.3 $\pm$ 0.4 &  10.0 $\pm$ 0.8  & 13.6 $\pm$ 1.0 & 26.7 $\pm$ \phm{1}2.1 &  22.5 $\pm$ \phm{1}1.6 & 516 & 704 \\
 NGC 419 &  5.38 $\pm$ 0.08 &  5.51 $\pm$ 0.08 &  5.94 $\pm$ 0.08 &  7.7 $\pm$ 2.9 &  6.0 $\pm$ 2.2 &  4.1 $\pm$ 1.5 &  20.6 $\pm$ 4.2  & 26.8 $\pm$ 5.5 & 53.4 $\pm$ 11.0       &  53.3 $\pm$ 11.0       & 560 & 799 \\
NGC 1651 &  4.91 $\pm$ 0.06 &  5.04 $\pm$ 0.06 &  5.48 $\pm$ 0.06 & 12.8 $\pm$ 1.0 &  9.9 $\pm$ 0.8 &  6.8 $\pm$ 0.5 &  10.2 $\pm$ 0.7  & 13.6 $\pm$ 1.3 & 27.3 $\pm$ \phm{1}2.7 &  20.4 $\pm$ \phm{1}2.5 & 315 & 584 \\
NGC 1718 &  4.83 $\pm$ 0.07 &  5.01 $\pm$ 0.07 &  5.43 $\pm$ 0.07 &  5.4 $\pm$ 0.6 &  4.2 $\pm$ 0.4 &  2.9 $\pm$ 0.3 &  13.0 $\pm$ 1.3  & 18.1 $\pm$ 1.8 & 35.5 $\pm$ \phm{1}3.5 &  27.8 $\pm$ \phm{1}2.7 & 406 & 650 \\
NGC 1751 &  4.81 $\pm$ 0.06 &  4.95 $\pm$ 0.06 &  5.38 $\pm$ 0.06 &  7.1 $\pm$ 0.9 &  5.6 $\pm$ 0.7 &  3.8 $\pm$ 0.5 &  10.9 $\pm$ 1.0  & 14.6 $\pm$ 1.4 & 29.0 $\pm$ \phm{1}2.7 &  25.4 $\pm$ \phm{1}2.4 & 353 & 509 \\
NGC 1783 &  5.42 $\pm$ 0.11 &  5.54 $\pm$ 0.11 &  5.98 $\pm$ 0.01 & 11.4 $\pm$ 2.2 &  8.9 $\pm$ 1.7 &  6.1 $\pm$ 1.2 &  17.6 $\pm$ 1.8  & 23.0 $\pm$ 2.4 & 46.0 $\pm$ \phm{1}4.8 &  39.9 $\pm$ \phm{1}4.2 & 403 & 584 \\
NGC 1806 &  5.10 $\pm$ 0.06 &  5.23 $\pm$ 0.06 &  5.66 $\pm$ 0.06 &  9.0 $\pm$ 1.2 &  7.0 $\pm$ 1.0 &  4.8 $\pm$ 0.7 &  13.7 $\pm$ 1.0  & 18.0 $\pm$ 1.4 & 35.9 $\pm$ \phm{1}2.7 &  31.4 $\pm$ \phm{1}2.4 & 370 & 613 \\
NGC 1846 &  5.24 $\pm$ 0.09 &  5.37 $\pm$ 0.09 &  5.80 $\pm$ 0.09 &  8.8 $\pm$ 0.7 &  6.8 $\pm$ 0.5 &  4.7 $\pm$ 0.4 &  16.3 $\pm$ 1.9  & 21.5 $\pm$ 2.6 & 42.9 $\pm$ \phm{1}5.1 &  35.8 $\pm$ \phm{1}4.6 & 567 & 757 \\
NGC 1852 &  4.66 $\pm$ 0.07 &  4.81 $\pm$ 0.07 &  5.24 $\pm$ 0.07 &  7.0 $\pm$ 0.8 &  5.5 $\pm$ 0.7 &  3.7 $\pm$ 0.4 &   9.4 $\pm$ 0.9  & 12.6 $\pm$ 1.2 & 24.9 $\pm$ \phm{1}2.4 &  23.7 $\pm$ \phm{1}2.2 & 312 & 432 \\
NGC 1987 &  4.74 $\pm$ 0.04 &  4.85 $\pm$ 0.04 &  5.26 $\pm$ 0.04 & 12.8 $\pm$ 3.0 & 10.1 $\pm$ 2.4 &  6.9 $\pm$ 1.6 &   8.7 $\pm$ 1.2  & 11.1 $\pm$ 1.8 & 21.7 $\pm$ \phm{1}3.6 &  20.4 $\pm$ \phm{1}3.4 & 234 & 424 \\
NGC 2108 &  4.71 $\pm$ 0.07 &  4.84 $\pm$ 0.07 &  5.24 $\pm$ 0.07 &  7.2 $\pm$ 0.8 &  5.7 $\pm$ 0.6 &  3.9 $\pm$ 0.4 &   9.8 $\pm$ 0.9  & 12.7 $\pm$ 1.2 & 24.5 $\pm$ \phm{1}2.3 &  21.3 $\pm$ \phm{1}2.0 & 230 & 359 \\
NGC 2154 &  4.61 $\pm$ 0.06 &  4.80 $\pm$ 0.06 &  5.21 $\pm$ 0.06 &  5.7 $\pm$ 0.5 &  4.4 $\pm$ 0.4 &  3.0 $\pm$ 0.3 &   9.9 $\pm$ 0.8  & 13.7 $\pm$ 1.1 & 26.9 $\pm$ \phm{1}2.1 &  23.6 $\pm$ \phm{1}1.9 & 431 & 625 \\
NGC 2173 &  4.67 $\pm$ 0.07 &  4.83 $\pm$ 0.07 &  5.26 $\pm$ 0.07 &  6.3 $\pm$ 1.1 &  4.9 $\pm$ 0.9 &  3.4 $\pm$ 0.6 &  11.4 $\pm$ 1.3  & 13.8 $\pm$ 1.7 & 27.2 $\pm$ \phm{1}3.3 &  22.7 $\pm$ \phm{1}3.5 & 431 & 589 \\
NGC 2203 &  4.95 $\pm$ 0.07 &  5.08 $\pm$ 0.07 &  5.51 $\pm$ 0.07 &  9.5 $\pm$ 1.6 &  7.4 $\pm$ 1.2 &  5.1 $\pm$ 0.8 &  11.2 $\pm$ 1.3  & 14.8 $\pm$ 1.7 & 29.4 $\pm$ \phm{1}3.4 &  25.5 $\pm$ \phm{1}2.9 & 475 & 652 \\
NGC 2213 &  4.46 $\pm$ 0.05 &  4.74 $\pm$ 0.05 &  5.13 $\pm$ 0.05 &  3.6 $\pm$ 0.3 &  2.8 $\pm$ 0.2 &  1.9 $\pm$ 0.2 &  10.4 $\pm$ 0.6  & 16.4 $\pm$ 1.0 & 30.8 $\pm$ \phm{1}1.9 &  20.2 $\pm$ \phm{1}1.2 & 329 & 502 \\
  LW 431 &  4.56 $\pm$ 0.07 &  4.68 $\pm$ 0.07 &  5.11 $\pm$ 0.07 &  9.1 $\pm$ 3.2 &  7.1 $\pm$ 2.5 &  4.9 $\pm$ 1.7 &   7.8 $\pm$ 1.5  & 10.0 $\pm$ 2.2 & 19.9 $\pm$ \phm{1}4.3 &  17.2 $\pm$ \phm{1}3.7 & 277 & 462 \\
 Hodge 2 &  4.70 $\pm$ 0.07 &  4.83 $\pm$ 0.07 &  5.25 $\pm$ 0.07 &  9.1 $\pm$ 2.3 &  7.2 $\pm$ 1.8 &  4.9 $\pm$ 1.3 &  10.2 $\pm$ 1.5  & 13.3 $\pm$ 2.4 & 26.1 $\pm$ \phm{1}4.7 &  19.3 $\pm$ \phm{1}3.5 & 363 & 520 \\
 Hodge 6 &  4.74 $\pm$ 0.07 &  4.94 $\pm$ 0.07 &  5.37 $\pm$ 0.07 &  5.5 $\pm$ 0.9 &  4.3 $\pm$ 0.7 &  2.9 $\pm$ 0.5 &  11.6 $\pm$ 1.3  & 16.6 $\pm$ 1.8 & 32.8 $\pm$ \phm{1}3.6 &  21.0 $\pm$ \phm{1}3.0 & \llap{$<$}238 & \llap{$<$}435 
\enddata
\tablecomments{Column (1): Name of star cluster. (2): Logarithm of adopted current
  cluster mass (in solar masses). (3): Logarithm of adopted cluster mass at
  an age of 10 Myr (no initial mass segregation case). (4): same as (3), 
  but for max.\ initial mass segregation case. 
  (5) Current cluster half-mass radius in pc. (6): Adopted  cluster half-mass radius 
   at an age of 10 Myr (no initial mass segregation case). (7): same as (6), 
  but for max.\ initial mass segregation case. 
  (8): Current central cluster escape velocity in \kms. (9): Central cluster escape 
  velocity at an age  of 10 Myr (no initial mass segregation case). (10): same as (8), 
  but for max.\ initial mass segregation case.  (11): same as (9), but for ``plausible'' 
  level of initial mass segregation (see \S\,\ref{sub:correl}). 
  (12): Value of {\it FWHM}$_{\it MSTO}$ in Myr. (13): Value of {\it W20}$_{\it MSTO}$ in Myr. 
  }
%\tablenotetext{a}{}
%  measurements of cluster and one background aperture.}
\end{deluxetable}

\end{turnpage}

\end{document}